\documentclass[%
preprint,
showpacs,
 amsmath,amssymb,
 aps,
prf,
]{revtex4-1}

\usepackage{graphicx}
\usepackage{dcolumn}
\usepackage{bm}
\usepackage{blindtext}
\usepackage{enumitem}
\usepackage{caption}
\usepackage{subcaption}
\usepackage{adjustbox}
\usepackage{booktabs}
\usepackage{array}
\usepackage[usenames, dvipsnames]{color}
\usepackage{amsmath}
\usepackage{amssymb}

\begin{document}


\title{Controlling droplet spreading with topography}

\author{P. Kant$^1$, A. L. Hazel$^1$, M. Dowling$^2$, A. B. Thompson$^1$ and A. Juel$^3$ }
\email{anne.juel@manchester.ac.uk}
\affiliation{$^1$School of Mathematics and Manchester Centre for Nonlinear Dynamics, University of Manchester, Oxford Road, Manchester M13 9PL, UK \\
$^2$Cambridge Display Technology, Technology Development Centre, Unit 12, Cardinal Business
Park, Godmanchester, Cambridgeshire, PE29 2XG, UK\\
$^3$Manchester Centre for Nonlinear Dynamics and School of Physics and Astronomy, University of Manchester, Oxford Road, Manchester M13 9PL, UK}

\date{\today}

\begin{abstract}
 We present a novel experimental system that can be used to study the
 dynamics of picolitre droplet spreading over substrates with
 topographic variations. We concentrate on spreading of a
 droplet within a recessed stadium-shaped pixel, with applications to the
 manufacture of POLED displays, and find that the 
 sloping side wall of the pixel can either locally enhance or hinder 
spreading depending on whether the topography gradient ahead of the contact line is positive or negative.
 Locally enhanced spreading occurs via the formation of thin pointed rivulets along the side walls of the pixel
 through a mechanism similar to capillary rise in sharp corners.

 We demonstrate that a thin-film model combined with an experimentally
 measured spreading law, which relates the speed of the contact line to
 the contact angle, provides excellent predictions of the
 evolving liquid morphologies. We also show that the spreading can be
 adequately described by a 
Cox--Voinov law for the majority of the
 evolution. The model does not include viscous
 effects and hence, the timescales for the propagation of the thin pointed rivulets are not captured.
Nonetheless,
 this simple model can be used very effectively to predict the areas
 covered by the liquid and may serve as a useful design tool for
 systems that require precise control of liquid on substrates.
 
\end{abstract}

\pacs{47.55.nd, 47.55.np, 47.55.dr}
\keywords{Spreading, topography, capillary rise}
\maketitle


\section{Introduction}
Controlled spreading of small amounts of liquid is a crucial
requirement in many applications such as microfluidic
(`lab-on-a-chip') devices \cite{stoneMicroFluidic,chin2007lab} and
inkjet-printing-based manufacture of displays
\cite{hallsCDT,inkjet_printingAccuracy,singh_inkjet,derby_inkjet}.
Several techniques including electro-wetting \cite{mugeleEW}, surface
energy patterning of substrates \cite{sirringhaus2000high} and
thermo-capillary pumping \cite{thermocapillary, kataoka} have been
proposed in the literature to drive small liquid volumes on surfaces.
In addition to these active methods, \citeauthor{PNAS} \cite{PNAS}
demonstrated that under certain circumstances, surface topography
such as grooves of rectangular cross-section can also be used to
passively control the spreading of liquid in the required regions. 
Geometric structures are used to generate a wide variety of
equilibrium liquid morphologies in biology \cite{herminghaus2008}, and
the morphology is strongly dependent on the size and
wettability of the geometric features.

The wide range of applications for spreading of fluid means that it has attracted considerable attention in the literature \cite{davis}.
The flow of a thin viscous film over topography typically induces depth inhomogeneities in the flowing film \cite{kalliadasis2000steady}. Hence, predictions of the propensity of a capillary-driven thin-film flow to smooth out grooves or holes on a substrate can serve as a design tool for lithography applications that require substrate levelling \cite{stillwagon1988fundamentals}. In the context of spreading on rough substrates, \citeauthor{shuttleworth} \cite{shuttleworth} demonstrated that topography modifies apparent contact angles, so that spreading can be locally enhanced or reduced compared with a flat substrate. Reduced spreading on a sharp-edged pillar was quantified by calculating of the excess fluid volume that could be accommodated in a sessile drop \cite{oliver1977resistance}.
More recently, substrates with a small periodic corrugation have been shown to yield multiple drop equilibria and generate stick-slip motion of the contact line in a two-dimensional model of a spreading droplet \cite{2dspreadingTopography} .

In the present paper our interest is in the spreading of a pico-litre
(pL) sized droplet on substrates with topography; a fundamental
process in the inkjet printing based manufacturing of POLED (Polymer
Organic Light Emitting Diode) displays, which usually involves the sequential deposition of several droplets. In order to manufacture high
resolution POLED displays, controlled spreading of the polymeric ink
is of the utmost importance and is usually achieved by patterning the
substrates using a photo-lithography method \cite{sirringhaus2000high}. These patterns (referred
to as pixels) are in the form of micron sized topographical features
or variations in the wettability of the
substrate or a combination of both.  Here we study the spreading of a
single droplet deposited in a pixel without wettability variations.

In the majority of previous studies of single-drop spreading over topographical features
\cite{PNAS, seemann2011, Oxlangmuir, science, ProcRoy.Soc., blobs} 
observations have been limited to the identification of the
final equilibrium morphologies. Recently \citeauthor{alice}
\cite{alice} considered fluid spreading resulting from the sequential deposition of partially overlapping droplets on flat substrates. They used a combination of high-speed micro-videography and simple mathematical
models, which allowed the time evolution of the liquid morphology to
be investigated. The present study follows that work by extending the
approach to include the effects of surface topography, but focuses on the spreading of a single deposited droplet.

In our new experimental setup (Fig. \ref{fig:schematic}) we use an
ink-jet printhead to deposit the droplet on an optically transparent
substrate, which allows visualization of the spreading of the single
drop from the bottom and the side.  We observe that the morphology of
the deposited liquid at equilibrium is dependent on the wettability of
substrate,
the slope of boundary wall of the pixel, and the initial
position of the droplet deposited in the pixel.
The wettability is characterised by the macroscopic
maximum equilibrium contact angle ($\theta_A$) of a droplet deposited
on  a flat region; and below a threshold value of $\theta_A$,
we find that the spreading is enhanced in the vicinity of the pixel
side walls through the growth of thin pointed rivulets. These are closely related to the classic phenomenon of capillary rise in polygonal geometries \cite{concusFinn, QuereUniversal, de1996mean,  concusMicroGravity}, which occurs when capillary-static surfaces in an inner corner become unbounded below a threshold value of the sum of corner and equilibrium contact angles \cite{concusFinn}. 

Models of the resulting capillary rise have focused on the pressure-driven rivulet flow in the corner \cite{radke,weislogel,weislogel2012compound}, and shown that resistance to flow decreases with a reduction in corner angle. However, these models do not consider the moving contact line at the tip of the rivulet.
 We find that the droplet's morphological evolution can be matched by a new theoretical
model over a wide range of contact angles and geometries.
The model assumes approximate quasi-static dynamics (shapes of constant
curvature in a thin-film approximation) driven by a contact-line spreading law $U(\theta)$,
the speed at which the contact line between the liquid and the
substrate advances as a function of the
contact angle. The exact spreading law can be measured experimentally, and, in principle,
should be measured for each different combination of substrate and
liquid; however, the
classic Cox--Voinov \cite{cox1986dynamics, voinov} law is found to provide a
reasonable approximation. The model does not include any viscous
effects which means that it is only in quantitative agreement with the
experiments at small times. At larger times, a given
morphology is attained more rapidly in the model compared to
experiments in which the viscous resistance retards the fluid motion.

The experimental apparatus and the nanofabrication of the substrates are described in \S \ref{sec:exp}, while the model and the numerical methods developed to predict droplet spreading are discussed in \S \ref{sec:model}. Experimentally measured spreading laws are presented in \S \ref{sec:spreading}, where they are used to validate the numerical model. In \S \ref{sec:results}, the effect  on spreading of the sign of the topography gradient ahead of the contact line is discussed first, followed by a detailed comparison between experimental spreading and model predictions for positive topography gradients, for which spreading is locally enhanced in the limit of small equilibrium contact angles. Conclusions are given in \S \ref{sec:conclusion}.

\section{Experimental methods} 
\label{sec:exp}

\subsection{Description of the experimental setup}

\begin{figure*}[ht!]
\centering
\includegraphics[clip, trim=2cm 3cm 0.5cm 1.5cm, width=1\textwidth]{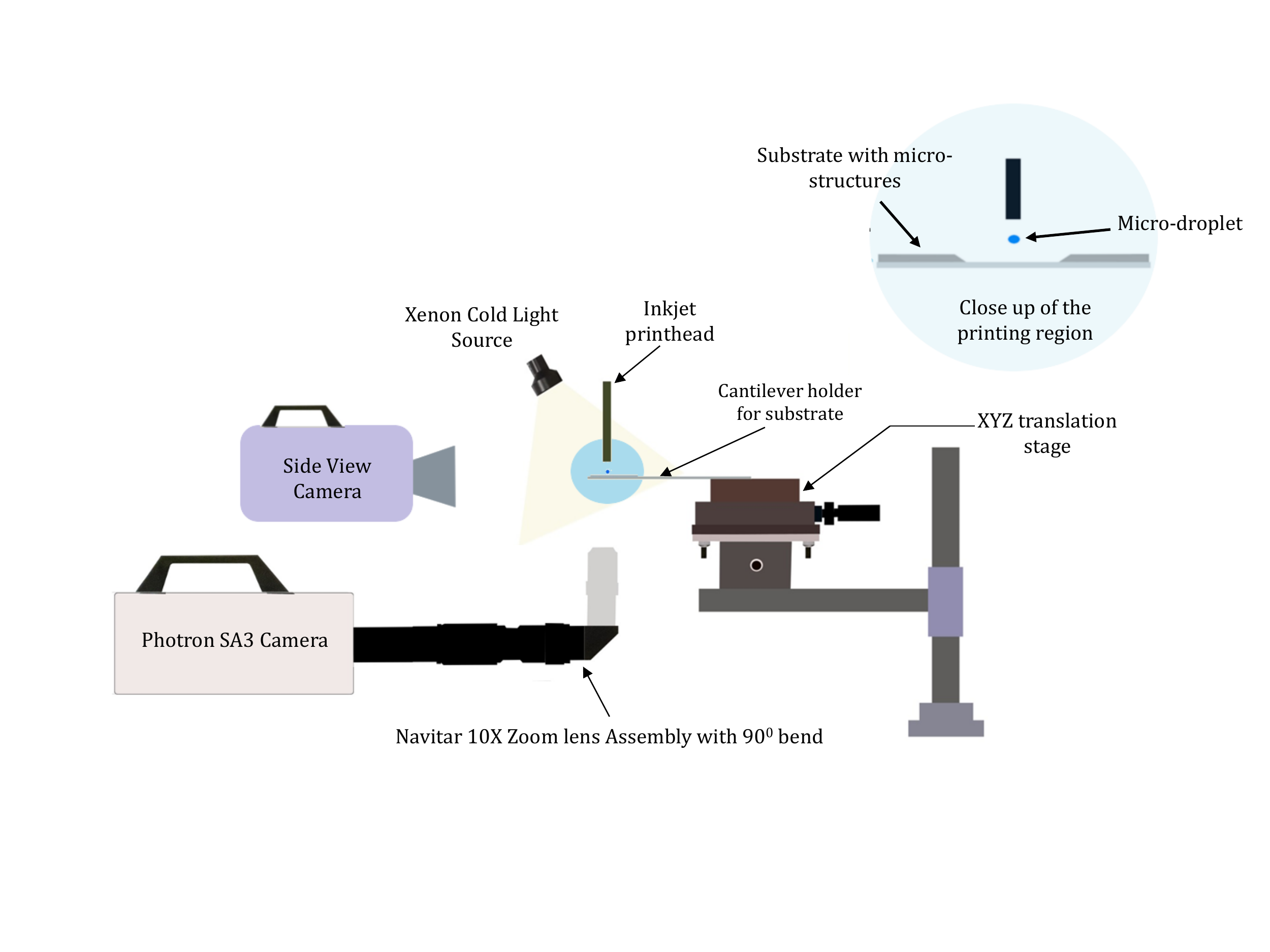}\\
\caption{Schematic diagram of the experimental setup in side view. A
  drop-on-demand inkjet printhead was used to deposit droplets on the
  horizontal substrate. The printhead was placed directly above the
  substrate at a distance of less than $1$~mm, as shown in the
  close-up. A high-speed camera fitted with long-distance magnifying
  (10x) optics recorded bottom views of the droplet spreading on the substrate. The early-time spreading of a droplet on the flat substrates was recorded using the same camera in side-view.}
\label{fig:schematic}
\end{figure*}

A schematic diagram of the experimental set-up is shown in Fig.
\ref{fig:schematic}. Droplets were deposited with an industrial grade
piezoelectric (drop-on-demand) ink-jet print-head (SX3, Fujifilm
Dimatix) positioned directly above the substrate at a distance of less
than $1$~mm. The print-head was powered by a waveform signal
generated with a NI-DAQ (6251, National Instruments) and amplified by
a voltage amplifier (PZD 350A, Trek Inc., USA). The diameter of the
print-head nozzle was 27~$\mu$m, and the droplet volume was kept fixed
at $V=$ 7.6 $\pm 0.4$~pL in all the experiments. The deposited fluid was a CDT proprietary solution used in POLED printing, with dynamic viscosity $\mu$ = 6.25 $\times$ 10$^{-3}$ ~Pa s, density $\rho$ = 1.066 $\times$ 10$^3$~kg m$^{-3}$ and surface tension $\sigma$ = 44 $\times$ 10$^{-3}$~N m$^{-1}$. 

The substrate (25~mm $\times$ 25~mm) was supported by a cantilevered holder, which comprised a square stainless steel frame with a thin outer lip to hold the substrate in position, and thereby enabled bottom view visualisation through the transparent substrate. The cantilevered holder was rigidly mounted on a screw-based motorized XY linear translation stage (Thorlabs PT1-Z8), which allowed positioning of the substrate with an absolute accuracy ($\xi_s$) of $\pm$5~$\mu$m. The entire assembly was placed on a ball-bearing vertical translation stage (M-MVN50, Newport) with manual positioning accuracy better than 1~$\mu$m. Vertical translation of the substrate allowed its positioning in the focal plane of the bottom view camera.

The small clearance between the print-head nozzle and substrate minimized the positioning error ($\xi_d$) in the placement of the droplets, caused by unavoidable disturbances in the ambient conditions, to $\pm$1.5~$\mu$m. 
The maximum error ($\xi$) in the placement of the droplet in the pixel is calculated as $\xi=(\xi_s^2 + \xi_d^2)^{1/2} = \pm 5.22~\mu $m \cite{inkjet_printingAccuracy}.
The vertical velocity of the drop with an in-flight radius of 10$\mu$m at impact was measured in earlier experiments \cite{alice}, using the same print-head, to be $V_i=2.2$~m/s. The Weber number associated with this impact  was low, \emph{We} = 2$\rho V_i^2 R_f / \sigma= 2.4$, and no splashing of the droplet at impact was observed. In the current experiments we used droplets of a slightly bigger size, with in-flight radius $R_f \sim \sqrt[3]{V/(4\pi/3)} \approx 12.2$~$\mu$m and we also did not observe any splashing at the time of impact.

Bottom views of the droplet spreading on substrates were recorded by a
high speed camera (Photron SA3, 128 $\times$ 336 pixel, 500 fps),
through a long distance zoom lens assembly (Navitar, with 10X
objective, Mitutoyo) with a 90$^\circ$ bend. The bottom view camera
was kept in a fixed position. The region of interest was illuminated
from the top by a cold Xenon lamp light (Xenon NOVA 300, Karl Storz).
In order to explore the early-time dynamics of the droplet deposition,
we also recorded side views  at 15,000 fps using similar magnifying
optics (without the 90$^\circ$ bend). Post-processing of the images
acquired by the camera was performed using standard built-in functions
of Matlab R2013a (Canny edge-detection algorithm). In order to extract
the details of the fluid footprint or profile from recorded grey-scale
images, the background was subtracted and the image contrast was
adjusted using \texttt{imadjust} functionality available in Matlab. The conversion from pixel to dimensional distance was achieved by imaging a micro-scale with divisions of 10~$\mu$m, which gave 1 pixel $\equiv 0.86$~$\mu$m for all bottom view images.

The setup was built on a breadboard (M-IG-35-2, Newport), which damped vibrations with frequencies ranging from $20--100$~Hz.
Additional vibration damping was achieved by placing the optical breadboard on a heavy steel table, and resting the feet of the table on sand. This reduced the maximum amplitude of vibrations to less than 1~$\mu$m. 
All the experiments were performed under ambient laboratory conditions, where the average evaporation rate per unit surface area of a sessile droplet of volume $7.6$~pl and contact angle $\sim$~50$^\circ$, was measured to be less than 0.168 $\mu$m/min ($<$ 5\% of total volume/min). The maximum duration of each experiment was less than 12~s except for the spreading law measurements, when the maximum duration was 1~min.

\begin{figure*}[ht!]
\centering
\includegraphics[clip, trim=2cm 15cm 2cm 6cm, width=1.0\textwidth]{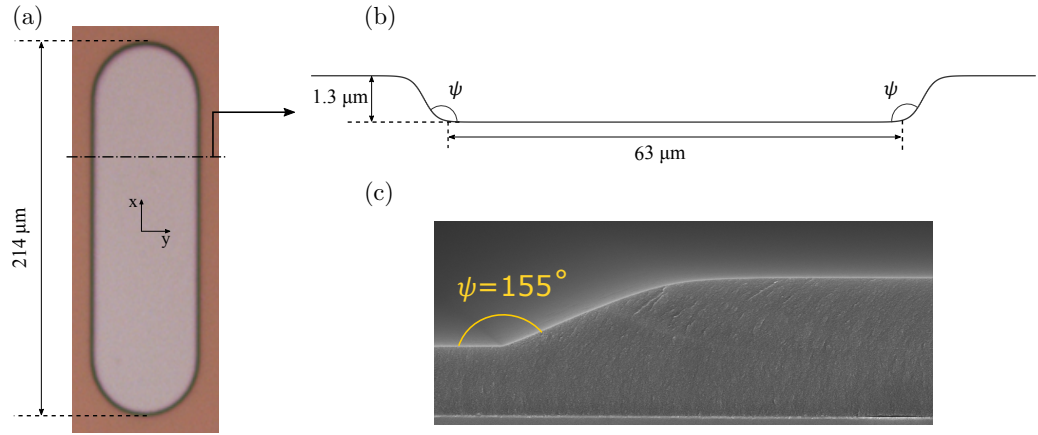}\\
\caption{Geometry of the pixel: (a) Top-view of a single stadium-shaped pixel. (b) Schematic diagram of a section of the pixel along the dot-dashed line shown in (a).  (c) Height profile of the side wall of a pixel image using Scanning Electron Microscopy. The angle between the flat bottom surface of the pixel and the side wall is $\psi = 155^{\circ}\pm1^{\circ}$ (corner angle). }
\label{fig:substrate}
\end{figure*}

\begin{figure*}[ht!]
\centering
\includegraphics[clip, trim=0.5cm 2cm 0.5cm 0.5cm, width=1.0\textwidth]{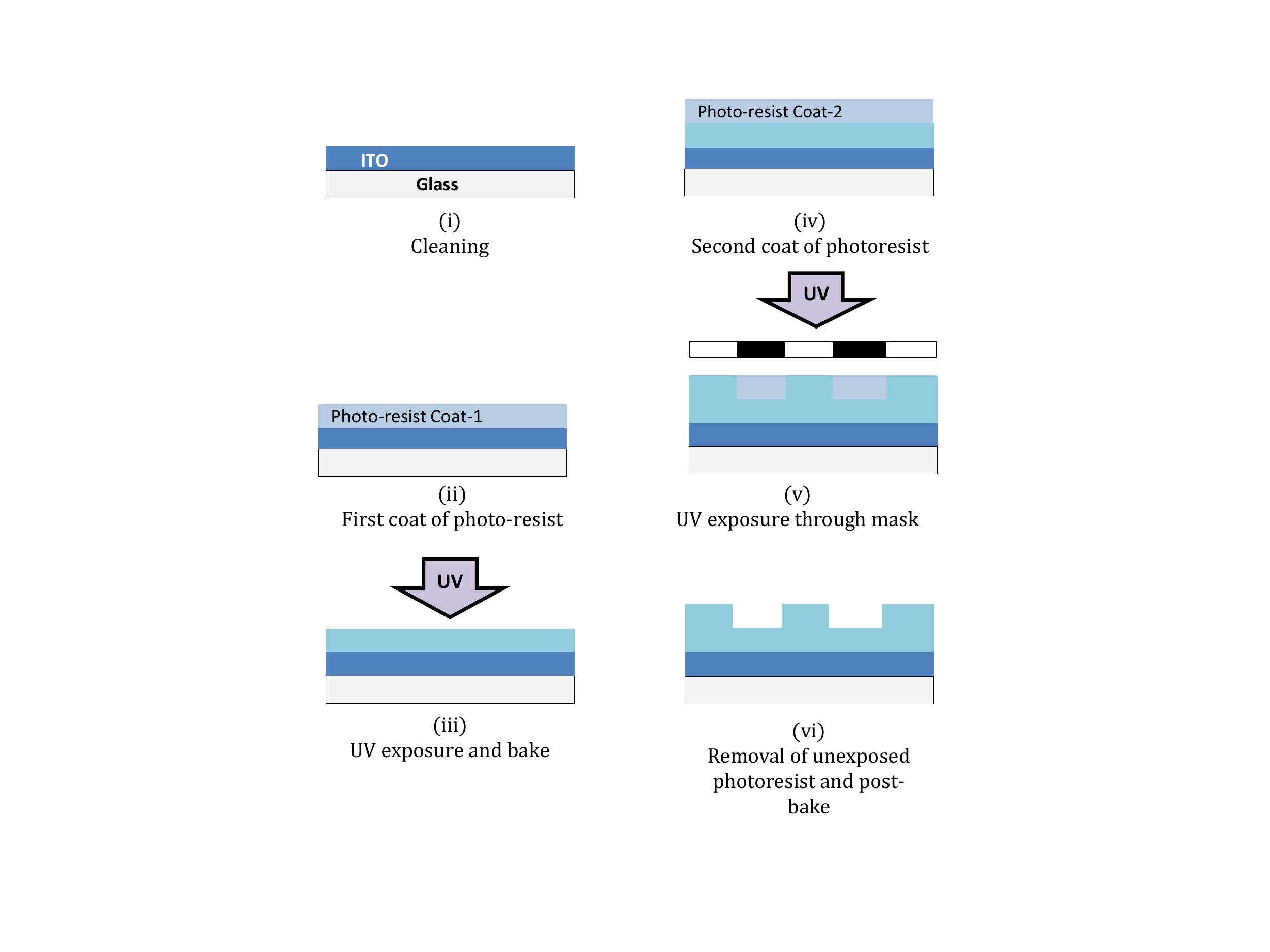}\\
\caption{Description of the photo-lithographic process used to make topographical features on glass substrates.}
\label{fig:substrate_preparation}
\end{figure*}

\subsection{Substrate preparation}

Droplet spreading was investigated in stadium-shaped wells
shown in Fig. \ref{fig:substrate}(a), which consisted of a rectangle
with elliptical ends. They will  be referred to hereafter as
pixels. The total length (width) of each pixel was 214~$\mu$m
(63~$\mu$m). A schematic diagram of the depth profile of the pixel is shown in Fig. \ref{fig:substrate}(b).
The side wall of each pixel was approximately
$D=1.3$~$\mu$m high and the height profile of the pixel boundary was
imaged using Scanning Electron Microscopy (see Fig.
\ref{fig:substrate}(c)). The angle between the flat bottom surface of
the pixel and its side wall was measured to be $\psi=155^\circ \pm
1^\circ$. This value was further confirmed by contact-stylus
profilometry (Dektak 150), within the accuracy of 1$^\circ$.

All substrates were manufactured in the clean room facility of \emph{Cambridge Display Technology} (CDT) using standard photo-lithographic  techniques as illustrated in Fig. \ref{fig:substrate_preparation}. 
Glass substrates were cleaned using a mixture of high-pressure deionized water and an ammonia based solution to remove particles, followed by several minutes treatment with UV light and ozone to remove organic contaminants. 
The substrates were subsequently coated with an approximately 1.3~$\mu$m thick layer of optically transparent negative-type photo-resist. This coated layer was then given a flood UV exposure with no photo-lithography mask in place, followed by a brief post-exposure bake to 110$^\circ$C and a 10 minute curing bake to 205$^\circ$C, both on hotplates. 
This treatment induced strong cross-linking of the photo-resist and provided a base to the pixel wells, using the same material and the same fabrication process as used for the surrounding banks, thus minimizing any contact angle differences between the two. 
The banks themselves were created from a second approximately 1.3~$\mu$m thick layer of the same photo-resist coated onto the substrate. This layer underwent exposure to UV, this time through a glass mask coated in a patterned Cr (chromium) layer which defined the pixel patterns, thus allowing the transfer of geometrical features onto the exposed area of the photo-resist. 
Following a brief post exposure bake to 110$^\circ$C on a hotplate, the unexposed areas of the photo-resist were removed using a commercial developer solution, and the resulting patterned region hardened through a curing bake at 205$^\circ$C for 10 minutes on a hotplate.


The wettability of the substrates was
quantified by the maximum equilibrium contact angle ($\theta_A$),
i.e. the contact angle at which a  droplet spreading on a flat region
of the substrate comes to rest. The CDT proprietary fluid exhibited
significant contact angle hysteresis on the substrate, with the minimum
equilibrium contact angle ($\theta_R$) being less than $1^{\circ}$. The wettability of the substrates was modified by aging following exposure to O$_2$ and Ar plasma (Minilab 060, Moorefield, UK) for 30 seconds. The contact angle $\theta_A$ measured immediately after the plasma treatment was reduced to a very low value ($\theta_A <1^\circ$), but increased progressively to $\theta_A = 35^{\circ}$ over a period of three weeks.

\section{Model and numerical methods}
\label{sec:model}

\begin{figure*}[ht!]
\centering
\includegraphics[clip, trim=2cm 13cm 2cm 9cm, width=1.0\textwidth]{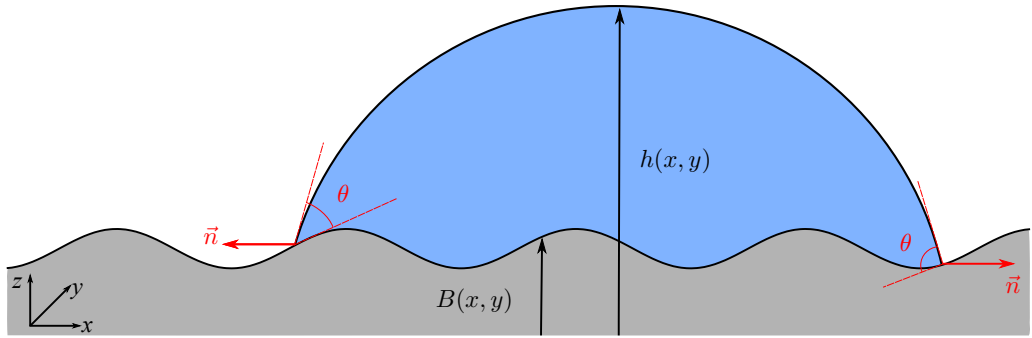}\\
\caption{Schematic diagram of the side-view of a droplet spreading on a surface with topography. The fluid lies in the region bounded by the interface $z=h(x,y,t)$ and the substrate $z=B(x,y)$.}
\label{fig:figure1}
\end{figure*}

Following \citeauthor{alice}'s \cite{alice} model of multiple droplets spreading on a flat substrate, we use thin-film equations to describe the drop-scale motion. We introduce Cartesian coordinates such that the incompressible fluid lies in the region $z=B(x,y)$ (the substrate) to $z=h(x,y,t)$ (the interface), as shown in Fig. \ref{fig:figure1}, and define the footprint $\Omega(t)$ as the projection of the  region where fluid and substrate meet onto $z=0$. The $x$ and $y$-axes are aligned along the length and width of the pixel, respectively. Throughout this paper, asterisks are used to distinguish non-dimensional quantities from their dimensional counterparts. The footprint radius $R_c$ of a droplet at rest on a flat surface with equilibrium contact angle $\theta_A$, and the capillary pressure $\sigma/ R_c$, are used to non-dimensionalise lengthscales and pressures, respectively. 

The characteristic length ($L$ = 100~$\mu$m) of the observed
morphologies is small, yielding a Bond number $Bo = \rho g L^{2} /\sigma \simeq 5 \times 10^{-3} \ll 1$, where $g$ denotes the acceleration of gravity, and  therefore we neglect any effects due to gravity.
Hence, in the limit of a static interface shape, the mean curvature $\kappa^*$ of the liquid-air interface is constant, and related to the excess internal pressure $P^*$ inside the liquid film through the Young-Laplace equation $ P^*(t^*) = \kappa^*(t^*) $. 
Under the thin-film approximation, the mean curvature is given by
\begin{equation}
\kappa^*(t^*) = -\nabla^{*2} h^*(x^*,y^*,t^*),  
\label{eq:kappa1}
\end{equation}
where $\nabla^*$ is the two-dimensional gradient operator. If the footprint $\Omega^*(x^*,y^*,t^*)$ and volume of the droplet $V^*$ are known at some instant, then we can determine $h^*(x^*,y^*,t^*)$ and the pressure $P^*(t^*)$, by solving the equations
\begin{eqnarray}
P^*(t^*) &=&  -\, \nabla^{*2} h^*(x^*,y^*,t^*)  \quad \mbox{for} \quad  (x^*,y^*) \in \Omega^* (t^*), \label{eq:kappa2} \\
V^* &=& \int_{\Omega^*(t^*)} \big( h^*(x^*,y^*,t^*) - B^*(x^*,y^*) \big) dx^* dy^*, \label{eq:vol}
\end{eqnarray}
with the boundary condition $h^* = B^*$ on the contact line $\partial \Omega^*(t^*)$.
As discussed in \S \ref{sec:exp}, the loss of liquid due to
evaporation is less than 5\%
of total volume  of a sessile drop per minute, and thus we neglect any loss of volume in the model.
The contact angle $\theta$ (in the small angle approximation) is determined from the relation
\begin{equation}
\theta = -\vec{n} \cdot \nabla^* (h^*-B^*),
\label{eq:theta}
\end{equation}
where $\theta$ is measured in radians, $\vec{n}$ is the two-dimensional unit normal directed out of
$\Omega^*$ (see Fig. \ref{fig:figure1}), and $h^*(x^*,y^*)$ is known from
the solution of Eq. (\ref{eq:vol}). 

\begin{figure*}[ht!]
\centering
\includegraphics[width=0.9\textwidth]{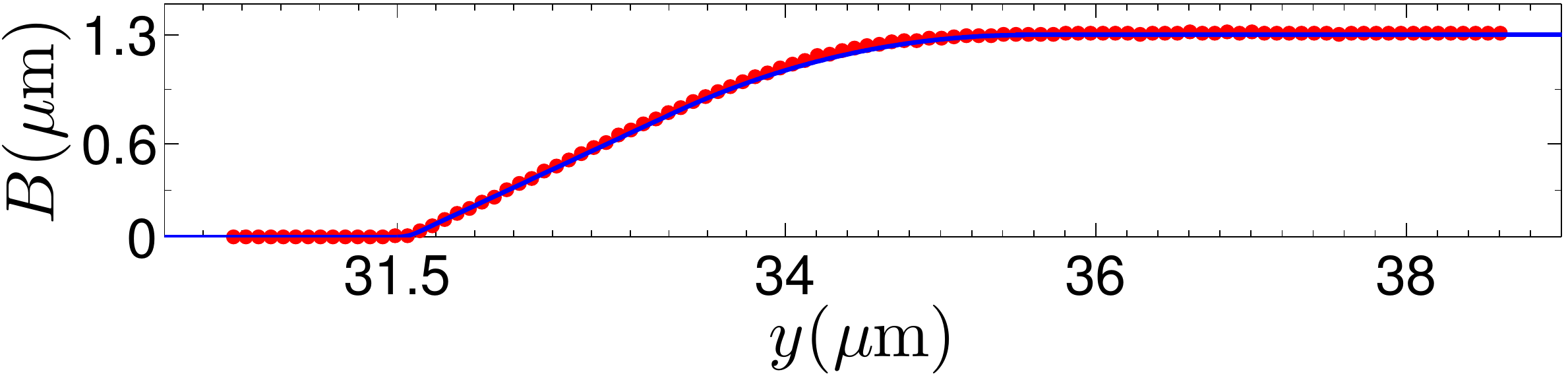}\\
\caption{Model of the side wall profile. The red symbols represent the experimentally measured height, while the blue line denotes the fit to these measurements used in the numerical calculations.  The convex and concave corners where the side wall joins the bottom and top horizontal surfaces of the well are approximated by arcs of circles.}
\label{fig:sidewall_shape}
\end{figure*}

The pixel geometry introduced in Fig. \ref{fig:substrate}  is
reproduced theoretically, by representing the side wall profile shown
in Fig. \ref{fig:substrate} with three linear segments smoothly connected by
two circular arcs, which closely fit the bottom and top corners of the side wall profile, as shown in Fig. \ref{fig:sidewall_shape}.

A kinematic condition enables the outward motion of the contact line with a normal velocity prescribed by the spreading law $U^*(\theta)$, so that
\begin{equation}
\vec{n} \cdot \frac{\mathrm{d}\vec{R^*}}{\mathrm{d}t^*} = U^*(\theta) \quad
\mbox{for} \quad \theta = - \vec{n} \cdot \nabla^* (h^*-B^*)
\label{eq:motion}
\end{equation}
where $\vec{R^*}(t^*)$ is the position of any material point on the
contact line $\partial \Omega^* (t^*)$. Hence, we solve Eq.
(\ref{eq:vol}) at each time step and update the position of the
contact line in accordance with Eq. (\ref{eq:motion}). Thus the entire
motion is controlled by the spreading law in combination with any
geometric changes. We
measure the spreading law for the advancing contact line
experimentally, see \S \ref{sec:spreading}, and we approximate the receding contact angle by zero,
so that the contact line is stationary if $\theta < \theta_A$.

For a given contact line position and fluid volume, the interface height
is determined by solving Eq. (\ref{eq:kappa2}), (\ref{eq:vol}) and the boundary condition $h^*=B^*$ on
the contact line. These equations are equivalent to solving a 2-D
Poisson equation in an arbitrarily shaped domain, and hence have a
unique solution. As described by \citeauthor{alice} \cite{alice}, we solve this
problem by using a finite element method with a boundary-fitted
triangulation of the fluid footprint, implemented using the open
source finite element library \texttt{oomph-lib} \cite{oomph}. We formulate the boundary
conditions for the Poisson problem so that the normal derivative of the
height weakly satisfies Eq. (\ref{eq:theta}), with the values of $\theta$ in this
condition determined so that the constraint $h^*=B^*$ is satisfied.

The only time derivative in the problem occurs in the advancing contact
line in Eq. (\ref{eq:motion}), which we discretise using the implicit BDF2 time-stepper.
Because the mesh is fitted to the fluid boundary, deformation of the 
contact
line requires deformation of the bulk mesh, which is treated by using a
``pseudo-elastic'' mesh: the mesh deforms as if it were an
elastic solid, and motion of the boundary contact line is accomplished
by applying normal stresses to the solid. As we will show in \S \ref{sec:results},
the fluid footprint and contact line can deform significantly (particularly
near sharp gradients in the substrate topography) resulting in a
poor-quality triangulation within the bulk mesh. We counter this by
re-triangulating the mesh every three timesteps, with area targets
for the new triangulation based on the spatial error
calculated by a Z2 error estimator \cite{Z2error} using the gradient of the
fluid height profile. The typical spatial discretisation consists of
approximately 5000 elements.
The numerical code is validated by calculating the spreading of a single
drop on a flat surface (see \S \ref{sec:spreading}). Numerical results for spreading
on topography are shown alongside the experimental results in \S \ref{sec:results}.


\section{Spreading laws and numerical validation}
\label{sec:spreading}

\begin{figure*}[ht!]
\begin{tabular}{cc}
\includegraphics[ width=0.47\textwidth]{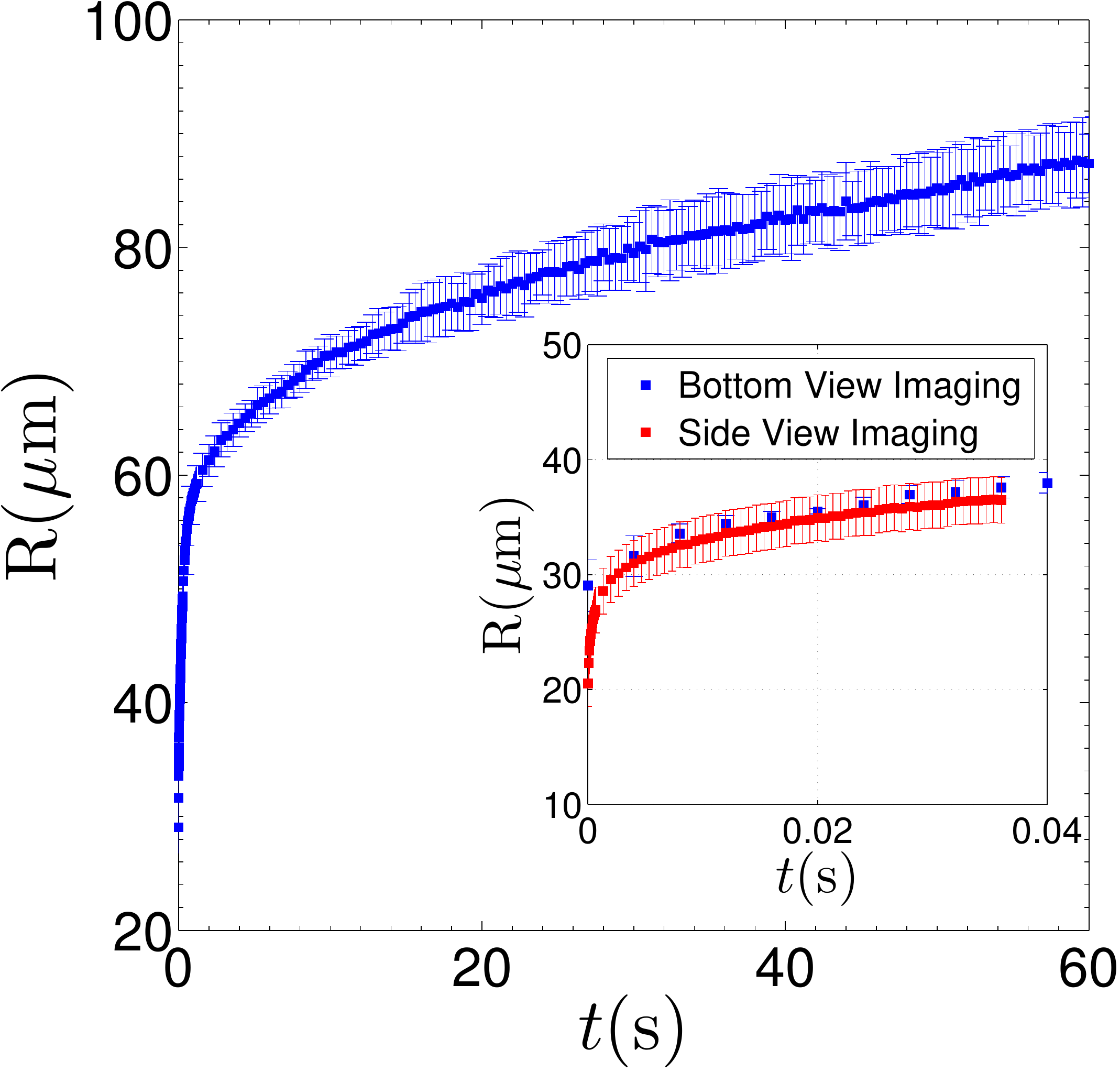} & \includegraphics[width=0.46\textwidth]{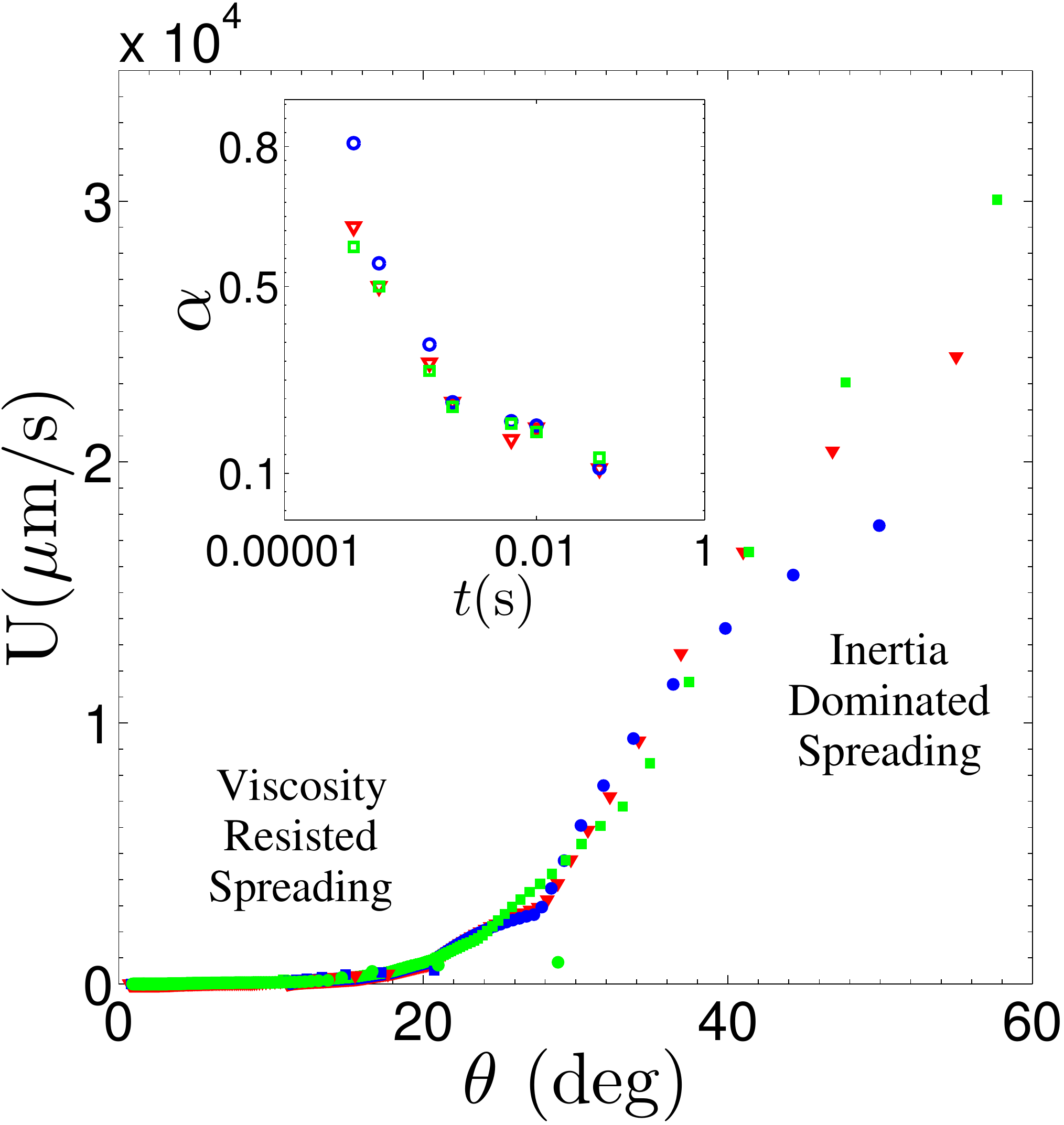}\\
(a)  & (b) \\[14pt]
\end{tabular}
\caption{Experimentally measured spreading of a droplet on a flat
  substrate ($\theta_A < 1^{\circ}$). (a) Evolution of the droplet
  radius with time. Inset: Early time dynamics recorded both in bottom
  view and side view visualizations. (b) Experimentally measured
  spreading law, each color corresponds to data from a single experiment. Inset: Time variation of the spreading pseudo-exponent,
  $\alpha$, defined in Eq. (\ref{eqn:exponent}): $\alpha \geq 0.5$  \cite{jaccoPOF, jaccoPRE} indicates inertially-resisted capillary spreading, and $\alpha = 0.1$ \cite{tanner} corresponds to viscosity-resisted capillary spreading (Tanner's regime).}
\label{fig:spreading_dynamics}
\end{figure*}

We measured spreading laws for two substrate wettabilities ($\theta_A
< 1^{\circ}$ and $\theta_A=11^{\circ}$)  by depositing a droplet on a
flat region of the substrate, and recording the position of its
contact line as a function of time. 
The radius of the  spreading droplet was determined in each frame by measuring  its perimeter from the bottom view images and fitting a circle to it using least squares. 
The evolution of the droplet
radius with time on the substrate with $\theta_A < 1^{\circ}$, obtained from bottom view images,  is shown in Fig. \ref{fig:spreading_dynamics}(a). 
Due to relatively low temporal resolution of the bottom view measurements, the fast spreading of the droplet shortly after its deposition was recorded in side view at 15,000 fps in separate experiments.
The inset image in Fig. \ref{fig:spreading_dynamics}(a) shows the evolution of the droplet radius just after deposition measured in side view (red squares) and bottom view (blue squares).
The error bars on the bottom view measurements in Fig. \ref{fig:spreading_dynamics}(a) indicate the standard deviation of the mean of three experiments. The error bars plotted on the mean of five experiments measured in side view indicate the maximum measurement error (2~$\mu$m) from image analysis.
The velocity of the contact line is calculated by differentiating a smooth cubic spline curve fitted through the experimental measurements of the radius as a function of time.  
The measurement of the dynamic contact angle from bottom-view images is based on the assumption that the droplet is a spherical cap of known volume at each instant in time.
The velocity of the contact line is shown as a function of the dynamic contact angle in Fig. \ref{fig:spreading_dynamics}(b).

In order to distinguish between different hydrodynamic spreading regimes, we follow \citeauthor{jaccoPOF} \cite{jaccoPOF}  and define the metric
\begin{equation}
\alpha =  \frac{\mathrm{d}( \ln R )}{\mathrm{d} (\ln t)},   \label{eqn:exponent}
\end{equation}
 where $\mathrm{d}$ denotes differentiation. In the limit where
 capillary spreading is resisted solely by either inertial or viscous
 forces \cite{jaccoPOF,tanner}, the metric $\alpha$ corresponds to a
 power law exponent, which takes constant values $\alpha \geq 0.5$
 \cite{jaccoPOF, jaccoPRE, bird}  and $\alpha =0.1$ \cite{tanner},
 respectively.
 The experimentally determined values of $\alpha$ decrease
 monotonically with time, see inset in Fig.
 \ref{fig:spreading_dynamics}(b), with viscosity-dominated spreading
 $\alpha \approx 0.1$ occurring at later times ($t> 10^{-2}s$).

\begin{figure*}[ht!]
\begin{tabular}{cc}
\includegraphics[width=0.46\textwidth]{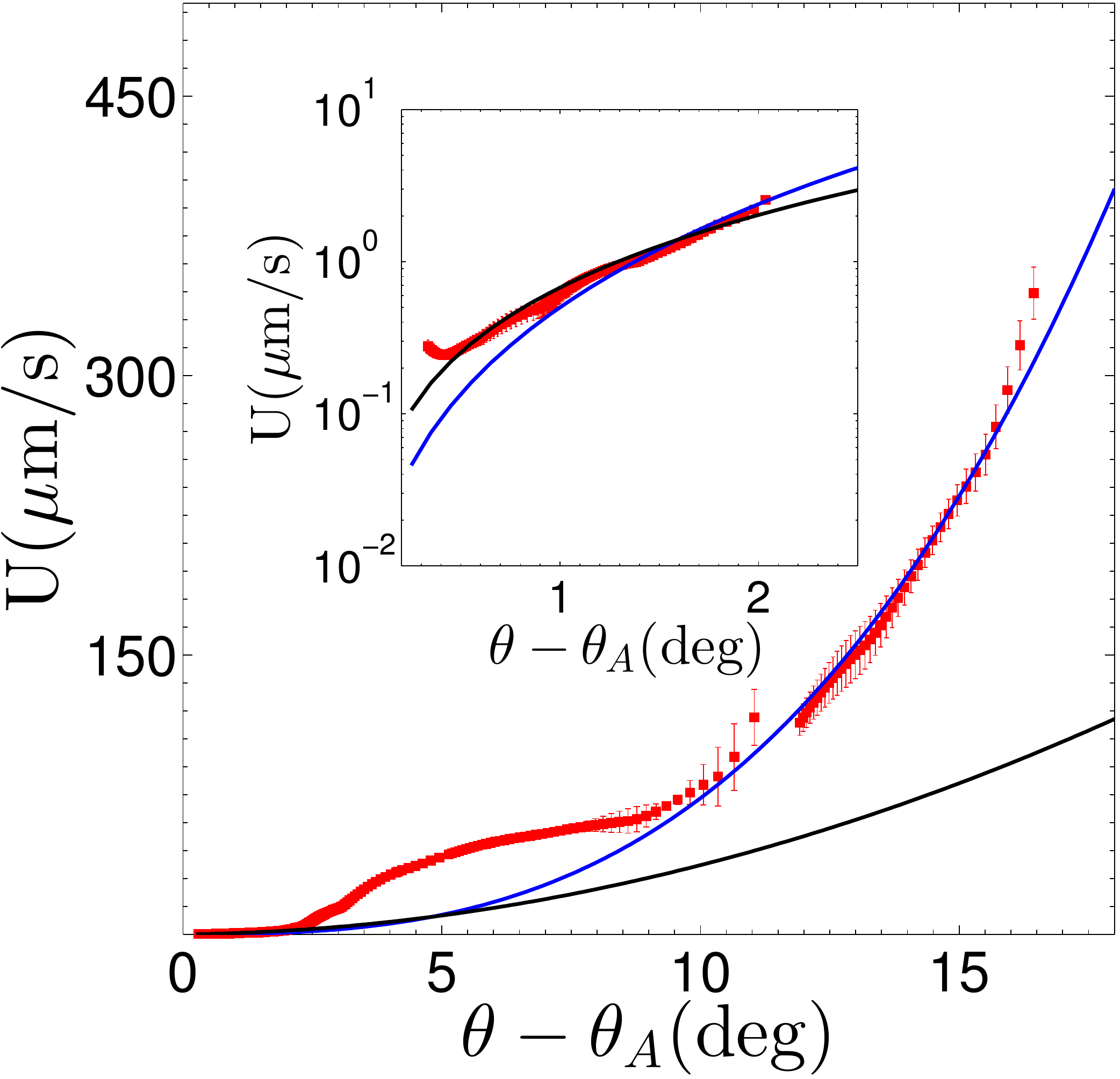} & \includegraphics[width=0.46\textwidth]{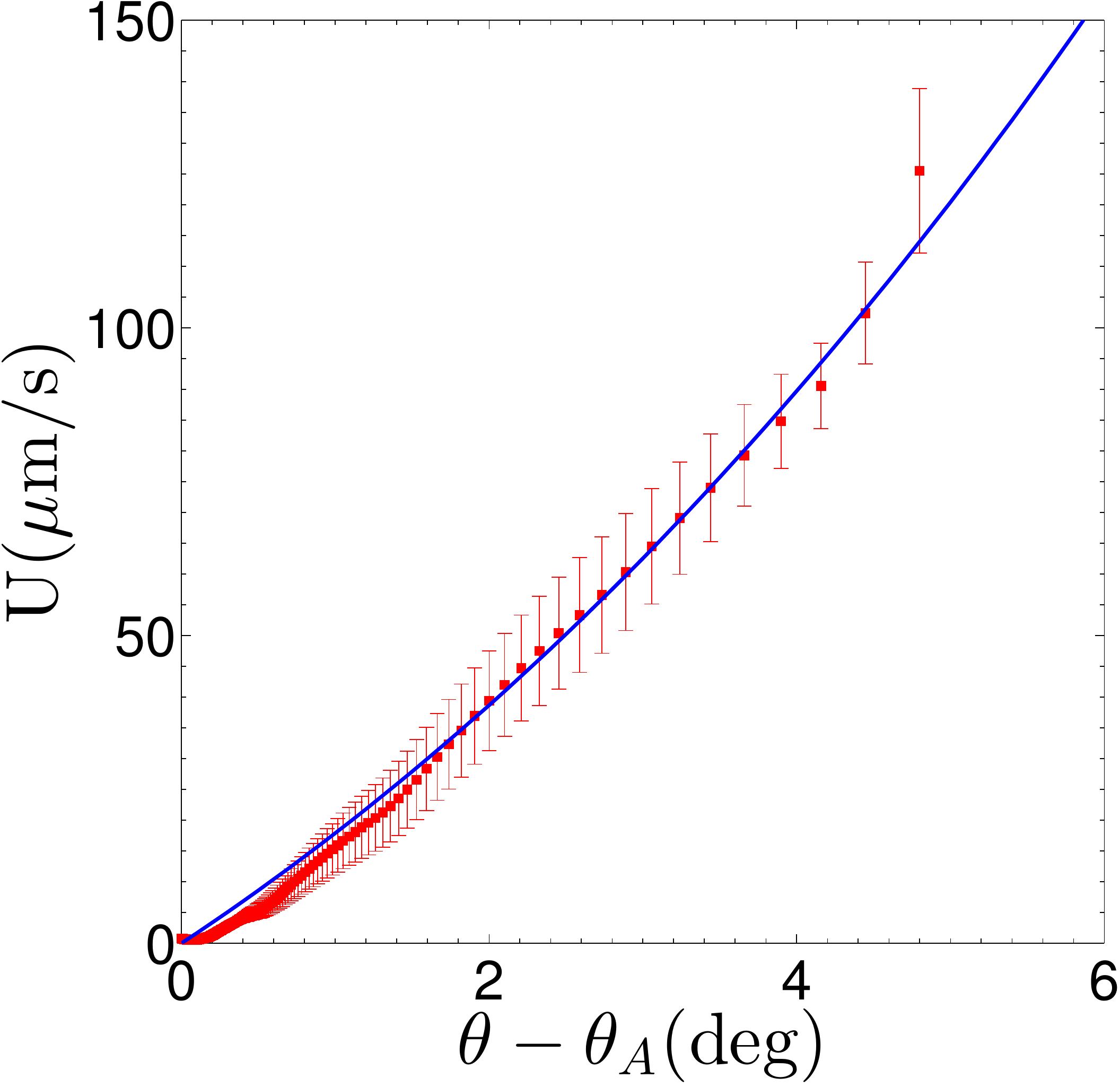}\\
(a)  & (b) \\[14pt]
\end{tabular}
\caption{Spreading laws measured on substrates with (a) $\theta_A < 1^{\circ}$  and (b) $\theta_A = 11^{\circ}$. The red squares represent the experimental data corresponding to the viscosity-dominated spreading regime. Blue and black lines represent the fits based on Cox--Voinov and de Gennes relations, respectively. (a) Inset: Comparison between the quality of the fit based on Cox--Voinov and de Gennes spreading laws at low dynamic contact angles.  The range of contact angles shown in (a) and (b) corresponds to the values required in the numerical simulations of spreading reported in \S \ref{sec:results}.}
\label{fig:fitted_spreadingLaw}
\end{figure*}

On the substrates with $\theta_A < 1^{\circ}$ and $\theta_A=11^{\circ}$, in the viscosity-dominated spreading regime ($\theta < 16^\circ$), we find that a reasonable approximation to the experimental data is provided by the Cox--Voinov spreading law \cite{cox1986dynamics, voinov}
\begin{equation}
U(\theta)=  U_1 (\theta^3 - \theta^{3}_A),
\label{eq:spreadinglaw}
\end{equation}
where $U_1$ is a velocity scale determined by a least-squares fit to  the experimental data.
The fits to the mean of the experimentally measured spreading laws are shown in Fig. \ref{fig:fitted_spreadingLaw}(a,b) for $\theta_A < 1^{\circ}$ and $\theta_A=11^{\circ}$, respectively.
The error bars on the experimental points indicate the standard deviation of the mean.
We use the coefficient of determination ($r^2$) \cite{goodnessOfFitting} to evaluate the goodness of fit.
In general, values of $r^2$ close to 1 indicate that the chosen model approximates the fitted data well, and $r^2$ = 1 corresponds to a perfect match between the model and experimental data.
On the substrate with $\theta_A=11^{\circ}$, the fit of the Cox--Voinov model yields $r^2$ = 0.994 (Fig. \ref{fig:fitted_spreadingLaw}(b)), suggesting a very good fit.
On the substrate with $\theta_A < 1^{\circ}$, the experimental data deviates from the Cox--Voinov model at low contact angles ($3^\circ < \theta - \theta_A < 9^\circ$, see Fig. \ref{fig:fitted_spreadingLaw}(a)), resulting in a lower value of $r^2$ = 0.922.
Moreover, at very low values of the dynamic contact angles $\theta < 2^{\circ}$, we found that the fit of the de Gennes spreading law \cite{de1986}
\begin{equation}
U(\theta)=  U_2 (\theta^2 - \theta^{2}_A),
\label{eq:spreadinglaw2}
\end{equation}
 is in better agreement ($r^2 = 0.991$) with the experiments (inset Fig. \ref{fig:fitted_spreadingLaw}(a)) compared to the fit of the Cox--Voinov spreading  law over the same range.

Based on these fits we get $U_1 = 6.321\times10^{-2}$~$\mu$m~s$^{-1}$~deg$^{-3}$ and
$3.632\times10^{-2}$~$\mu$m~s$^{-1}$~deg$^{-3}$ for the substrates with
$\theta_A < 1^{\circ}$ and $\theta_A=11^{\circ}$, respectively, which
are similar to the value reported in \citeauthor{alice} \cite{alice}
($U_0 = 5.31 \times 10^{-2}$~$\mu$m~s$^{-1}$~deg$^{-3}$),
for a similar liquid spreading on ITO coated glass substrates.
The fit of the de Gennes spreading law to the experimental measurements at low contact angles yields the velocity scale $U_2 = 31.771 \times10^{-2}$~$\mu$m~s$^{-1}$~deg$^{-2}$.

\begin{figure*}[ht!]
\centering
\includegraphics[width=0.5\textwidth]{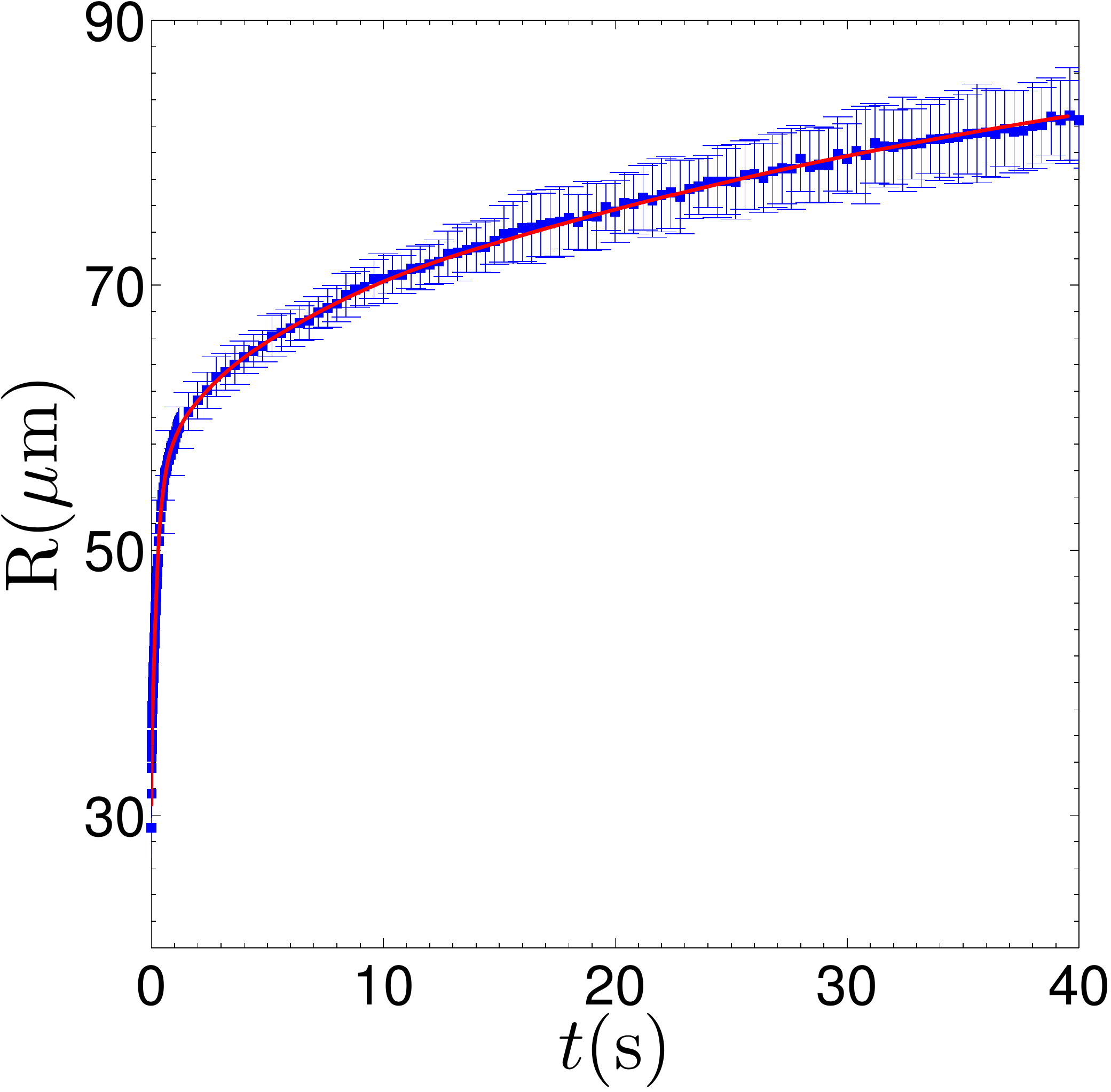}\\
\caption{Validation of the numerical model: comparison between numerical computations (red line) and experiments (blue squares) for the dependence of droplet radius on time with $\theta_A < 1^{\circ}$. The input to the simulations was an average spreading law calculated from the experimental data shown in Fig. \ref{fig:spreading_dynamics}(a).}
\label{fig:validation}
\end{figure*}

Finally, we used the experimentally measured spreading law to confirm the
consistency of the numerical model, in the case of a droplet spreading
on a flat substrate. The results of numerical calculations are
compared in Fig. \ref{fig:validation} with the experimentally
measured droplet radius as a function of time, which was originally
used to determine the spreading law. The prediction from the model is
in quantitative agreement with the average of three experiments to
within error bars. Hence, the model is able to correctly predict both
the fast spreading associated with the impact-driven dynamics at early
times, and the subsequent slowing down, as one would hope.

\section{Results}
\label{sec:results}

We shall begin with the discussion of the role of the sign of any gradients in topography on
the spreading of droplets. We find that droplets on a raised
topography in which the contact line encounters a downhill slope are
restricted in their spreading. Conversely, the spreading of
droplets within depressions (pixels) is locally enhanced where the contact line encounters an uphill slope at the pixel boundary.
We next present detailed comparisons between the numerical predictions
using the model discussed in \S \ref{sec:model} and experiments
for the droplet spreading inside a pixel with $\theta_A < 1^{\circ}$ and $\theta_A = 11^{\circ}$.
Finally, we conclude with a discussion on the effect of corner
smoothness on the spreading enhancement of the liquid within pixels.

\subsection{The effects of the sign of the topography gradient}

\begin{figure*}
\centering
\includegraphics[trim=3cm 12cm 3cm 10cm, clip=true, width=0.9\textwidth]{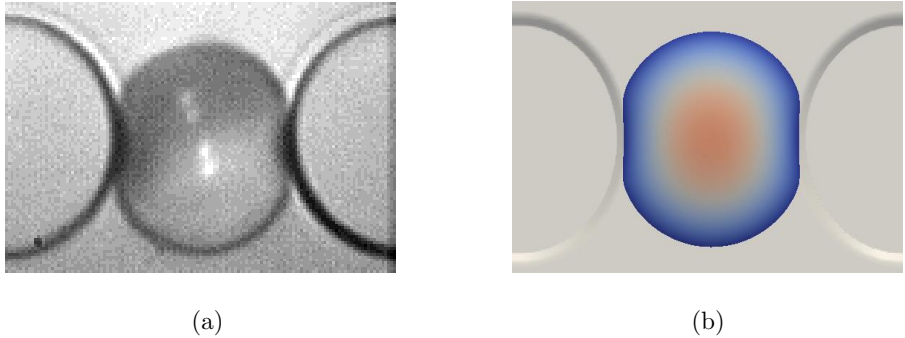}
\caption{(a) Snapshot of a droplet after spreading on the raised flat surface with $\theta_A=11^\circ$, which separates the ends of two pixels. (b) Numerical prediction corresponding to the configuration described in (a). The colour scheme indicates the height of the fluid interface relative to the substrate, which is maximum in the centre of the droplet's footprint.}
\label{fig:ConcaveConvex}
\end{figure*}

The expression for the contact angle (\ref{eq:theta}) shows that
the additional term due to topography gradients is $\vec{n}\cdot \nabla
B(x,y)$. Thus, for the same fluid height,  the contact angle is enhanced (reduced) when $\vec{n}\cdot \nabla
B(x,y) > 0$ ($\vec{n}\cdot \nabla B(x,y) < 0$), which corresponds to the contact line encountering a
surface sloping uphill (downhill) \cite{shuttleworth}. As shown in Fig. \ref{fig:ConcaveConvex}, a droplet positioned on the elevated flat surface between two pixels is constrained to remain there, and thus,
it spreads by conforming to the outline of the flat surface  until the
condition $\theta = \theta_A$ is reached. Hence, topography provides
an effective means of restricting small volumes of liquid to a
specific region of the substrate. The remainder of the paper will
focus on droplets spreading inside a pixel where the contact line
encounters an uphill slope, in order to characterise the local spreading enhancement via thin rivulet formation along the side walls.

\subsection{Rivulet formation inside a pixel}
\label{sec:resultsB}

\begin{figure*}[ht!]
(a) \hspace{4cm} (b) \hspace{4cm} (c)
\centering \includegraphics[trim=2.5cm 9cm 2.5cm 6cm, clip=true, width=1.0\textwidth]{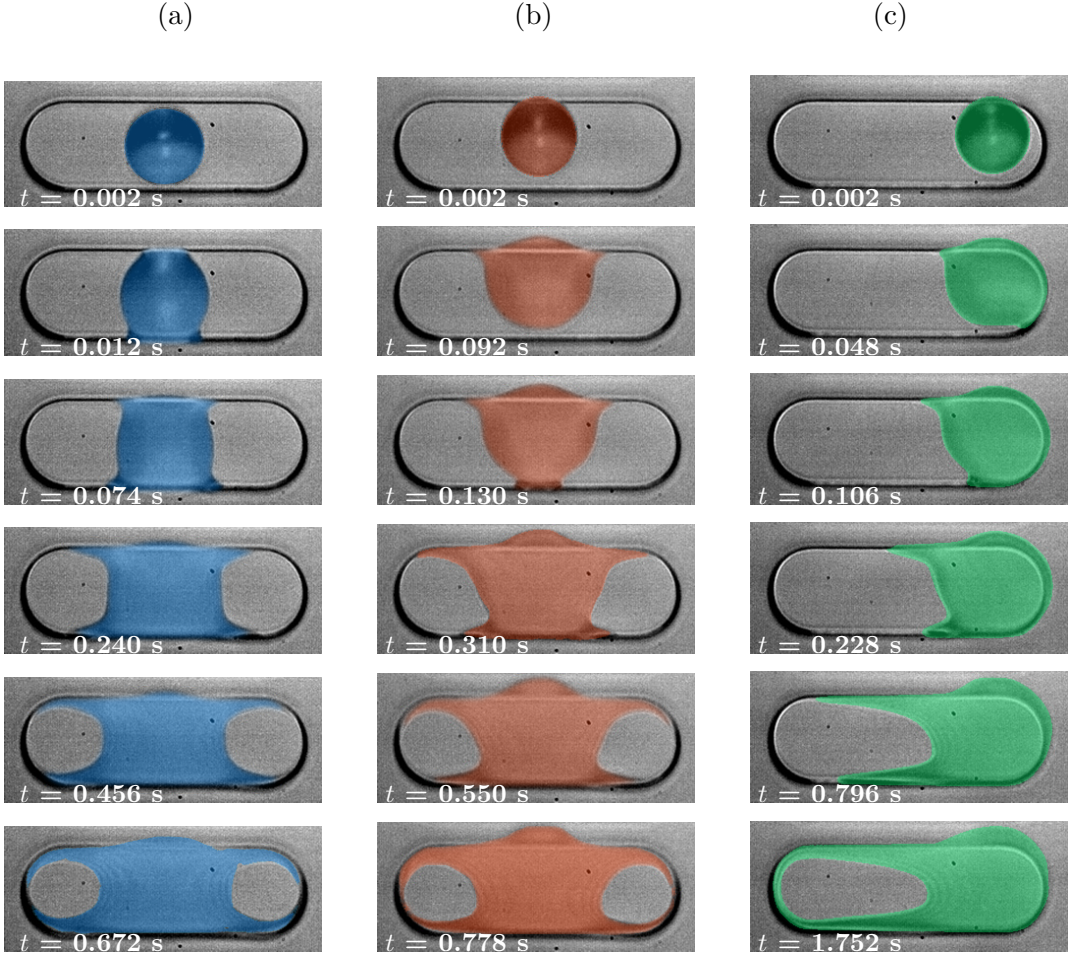}\\
\caption{Spreading of a droplet deposited inside a pixel from an
  initially circular footprint to a thinning film with rivulets
  growing along the side walls. (a), (b) and (c) show spreading for
  three different locations of droplet deposition within the pixel. The
  equilibrium contact angle is $\theta_A < 1^{\circ}$. Note that the
  last image of each sequence does not correspond to the final shape of the liquid film. The footprints of the liquid films recorded at the end of the experiment ($t \simeq 10$~s) are shown in Fig. \ref{fig:variation_with_theta}.}
\label{fig:filaments_0p5Deg}
\end{figure*}

Droplet spreading experiments were performed for five values of the
equilibrium contact angle, $\theta_A < 1^\circ$, and
$\theta_A=7^\circ, \; 11^\circ,\; 18^\circ, \; 23^\circ$.
Examples are shown in Fig.
\ref{fig:filaments_0p5Deg} of the evolution of the liquid footprint
for $\theta_A < 1^\circ$ and three different initial deposition locations
within the pixel. In all cases, the droplet spreads from an initial approximate
hemisphere at $t=0$~s into an elongated shape with thin rivulets
that travel along the pixel boundary before meeting and coalescing to
leave one or two holes within the liquid footprint. 
Thus, for strongly wetting
pixels, the spreading morphology is strongly influenced
by the presence of the side-wall. For a droplet deposited in the centre
of the pixel, as shown in Fig. \ref{fig:filaments_0p5Deg}(a),
spreading takes place predominantly inside the well, symmetrically
about both the vertical and horizontal symmetry planes, with minimal spillage
over the side walls. Droplets deposited close to the
side wall of the pixel, as in Fig. \ref{fig:filaments_0p5Deg}(b)
and \ref{fig:filaments_0p5Deg}(c), deform asymmetrically,
and their spreading is associated with significant overspill.
On these strongly wetting substrates, as the liquid morphology evolves beyond the times shown in Fig. \ref{fig:filaments_0p5Deg}, the surface area of the liquid footprint increases and the remaining holes reduce in size. This leads to a significantly enhanced volumetric evaporation rate of 2.521~pl/min ($\approx$ 30\% of the deposited volume/min) at $t=10$~s. Hence, only the experimental observations corresponding to $t< 10$~s are reported on these substrates.

\begin{figure*}[ht!]
\hspace{-1cm} (a) \hspace{3.5cm} (b)
\centering
\includegraphics[trim=2.5cm 9cm 2.5cm 6cm, clip=true, width=0.9\textwidth]{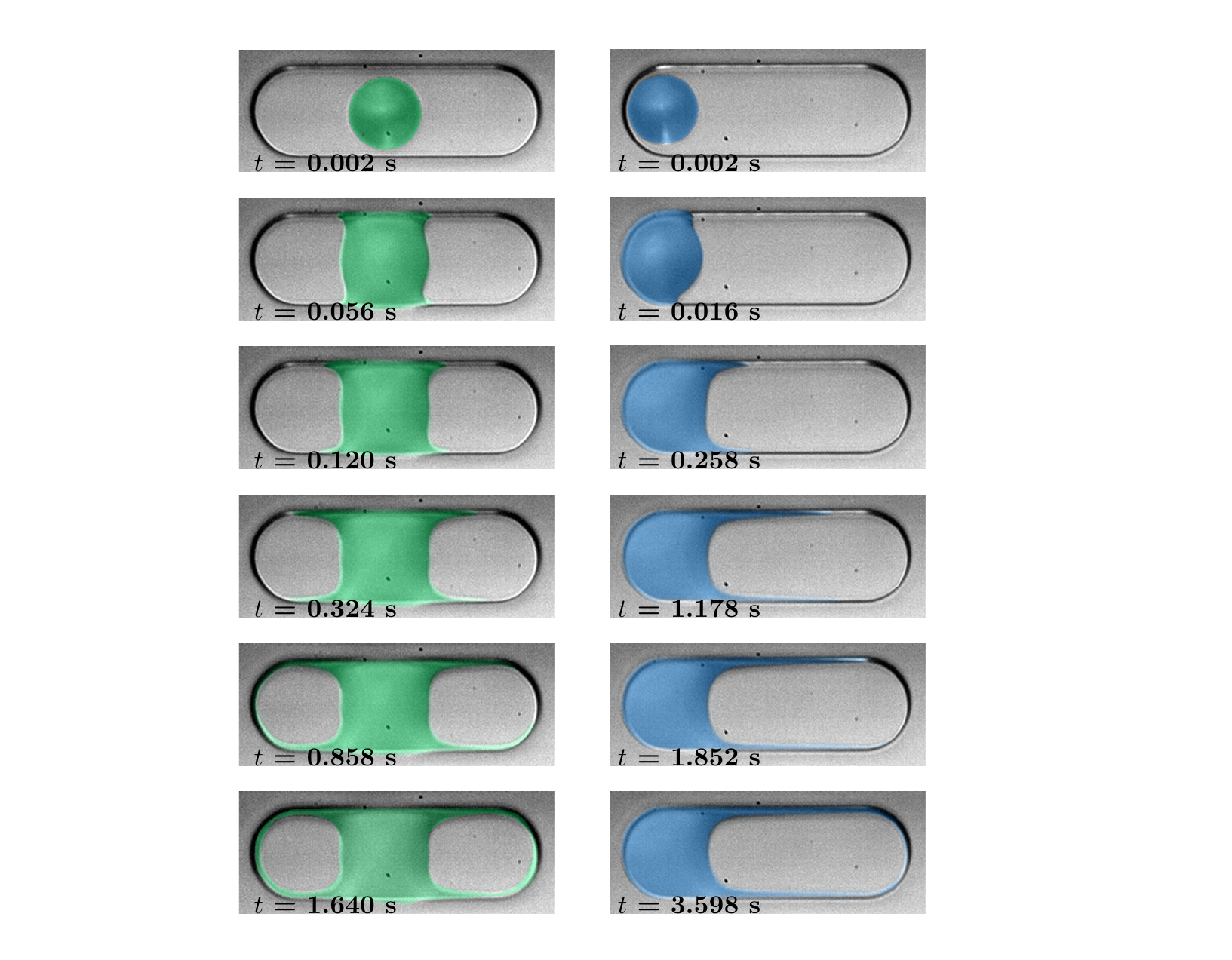}\\
\caption{Spreading of a droplet deposited inside a pixel from an
  initially circular footprint to a thinning film with rivulets
  growing along the side walls. (a) and (b) show spreading for two
  different locations of droplet deposition within the pixel. The
  equilibrium contact angle is $\theta_A= 11^{\circ}$. Note that the
  last image of each sequence does not correspond to the final shape of the liquid film. The footprints of the liquid films recorded at the end of the experiment ($t\simeq 10$~s) are shown in Fig. \ref{fig:variation_with_theta}.}
\label{fig:filaments_11Deg}
\end{figure*}

 Figure \ref{fig:filaments_11Deg} shows typical spreading in pixels with
 $\theta_A =11^{\circ}$. The footprint of the spreading drop shows a
 similar evolution to that shown in Fig. \ref{fig:filaments_0p5Deg},
 which depends on the initial deposition location. However, the
 rivulets formed along the side walls of the pixel are thinner for this
 higher contact angle and the
 surface covered by the fluid within the pixel when the rivulets
 coalesce is much reduced, as would be expected because the substrate
 is less wetting.

\begin{figure*}[ht!]
\centering \includegraphics[trim=2.8cm 9cm 1.5cm 12cm, clip=true, width=0.9\textwidth]{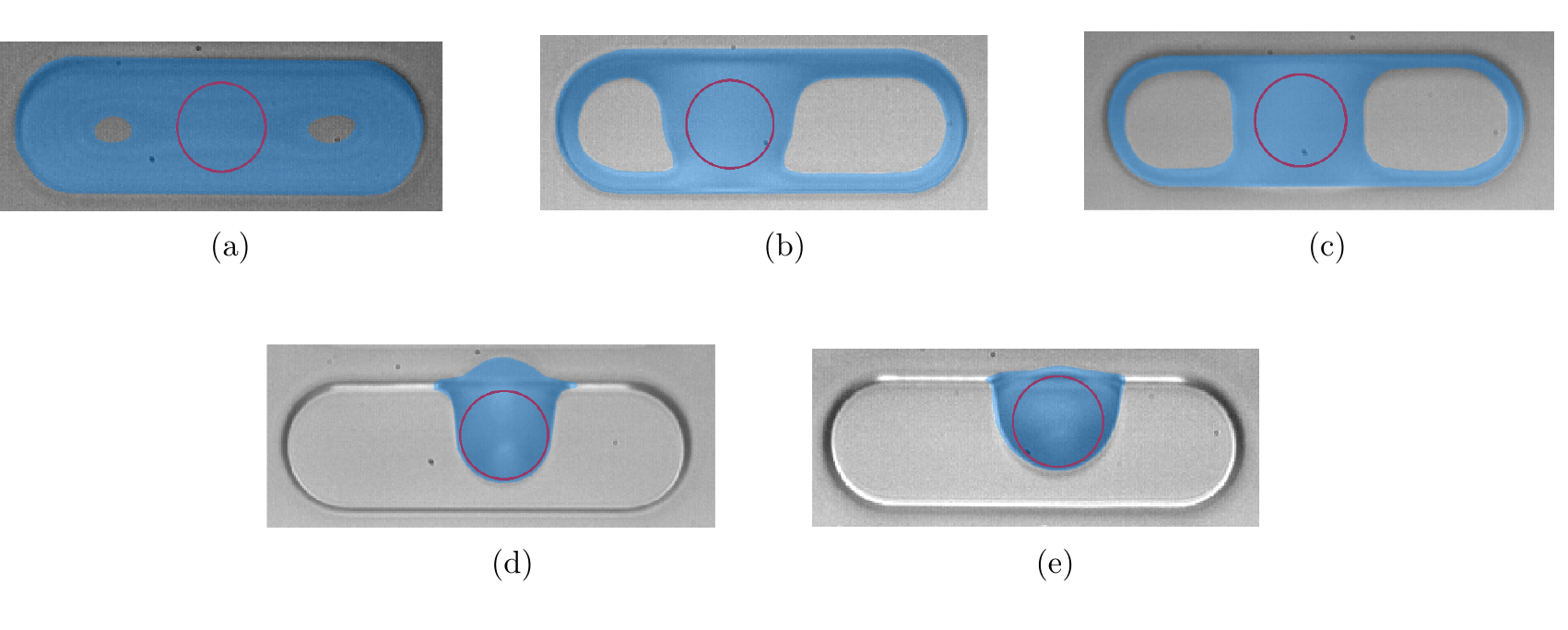}\\
\caption{Liquid footprints of the droplets deposited inside a
  pixel for different values of the equilibrium contact angle  (a)
  $\theta_A <1^{\circ}$, (b) $\theta_A=7^{\circ}$, (c) $11^{\circ}$,
  (d) $18^{\circ}$, and (e) $23^{\circ}$. Rivulets cease to form along
  the pixel boundary for $\theta_A > 11^{\circ}$. The blue shaded regions indicate the area covered with fluid, and the red circles mark the position of the droplet at deposition. For $\theta_A < 1^\circ$, fluid spreads over a larger proportion of the pixel surface, so that the volumetric evaporation rate is significantly increased compared with a sessile drop. As discussed in the text, we restrict our experimental observations to $t<10$~s, where evaporation remains limited.  On the less wetting substrates, evaporation does not become significant within the first minute of spreading, and the morphologies shown are quasi-equilibria.}
\label{fig:variation_with_theta}
\end{figure*}

Figure \ref{fig:variation_with_theta} shows the final experimental liquid footprint shapes (recorded at $t=10$~s) for five values of the equilibrium contact angle, in the case of a centrally placed droplet.  Rivulets cease to form for contact angles
 $\theta_A > 11^\circ$.  Instead, the footprint of the
 droplet spreads very little and develops a liquid morphology sometimes
 referred to as a `blob' \cite{blobs}. For all the substrates
 investigated, we performed a minimum of five experiments to check the
 robustness of the spreading phenomena. The general morphologies were
 identical in all repeated experiments and the final areas occupied by
 the liquid differed by less than 7\%.  

Our experimental results are consistent with theoretical predictions
by \citeauthor{concusFinn} \cite{concusFinn} of a capillary free
surface in a tube of polygonal cross-section, in the absence of
gravity. They found that a free surface
with constant mean curvature in a wedge of angle $\psi$ that meets
the walls with contact angle $\theta_A$ is unbounded in the corners when
\begin{equation}
\frac{\psi}{2} + \theta_A  < 90^\circ. \label{eqn:concusandfinn}
\end{equation}
The substrates used in
the experiments have a wedge angle (between the side wall and inner
part of pixel) of $\psi = 155^{\circ}$.  The threshold value of
$\theta_A$ calculated from the above relation is 12.5$^{\circ}$. This
threshold is above  $\theta_A=11^\circ$, the largest value of $\theta_A$ for
which rivulets were observed, but below $\theta_{A}={18}^{\circ}$,
above which rivulets are not observed ($\theta_A = {18}^{\circ}$ is the next available experimental data point after $11^{\circ}$).
These findings suggest that the growth of
rivulets in our experimental configuration is driven by capillary
forces, and is analogous to capillary rise in a wedge.

\subsection{Comparison between experiments and numerical predictions}

\begin{figure*}[ht!]
\centering
\includegraphics[trim=4cm 14cm 4cm 2.5cm, clip=true, width=0.75\textwidth]{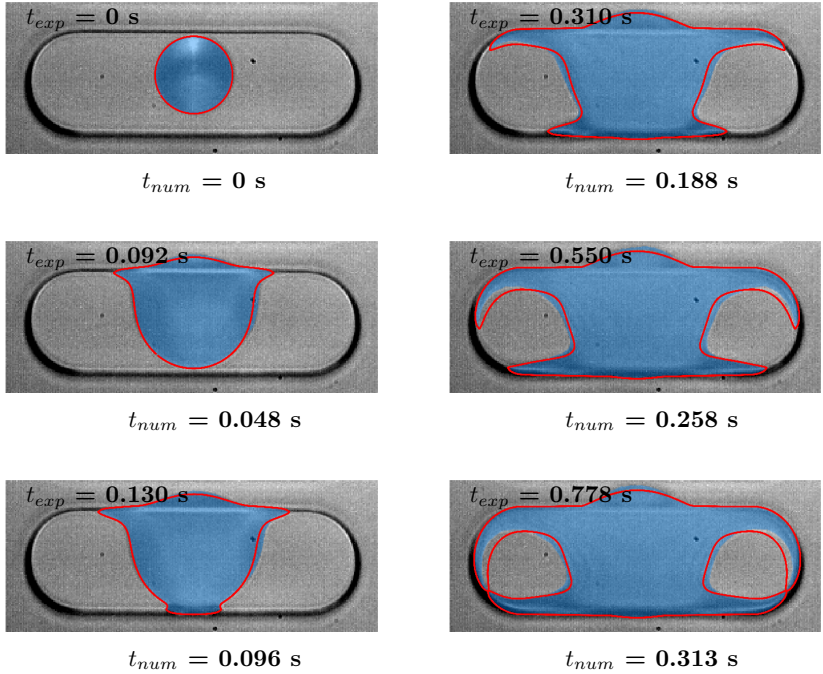}\\
\caption{Comparison between numerical calculations and experimental
  measurements for a series of snapshots of a spreading droplet
  deposited inside a pixel. Blue shaded regions correspond to the
  experimentally measured footprint (\textbf{$\theta_A <
    1^{\circ}$}). The red lines show the outline of the droplet
  footprint calculated numerically using the experimentally measured
  spreading law presented in Fig. \ref{fig:spreading_dynamics}(b).} 
\label{fig:comparison_0p5}
\end{figure*}

A comparison is shown in Fig. \ref{fig:comparison_0p5} between
numerical and experimental snapshots of the evolving footprint in a
pixel well with $\theta_A < 1^\circ$  for identical deposition
locations. The last numerical snapshot corresponds to the instant when
the growing rivulets first intersect  at both ends of the pixel,
beyond which the simulations were terminated. We used the
experimentally measured spreading law shown in Fig.
\ref{fig:spreading_dynamics}(b) to advance the contact line in the
computations, as discussed in \S \ref{sec:model}. The model
predicts the formation and growth of liquid rivulets along the side
walls of the pixel. 
To compare computations with the experimental snapshots of the evolving footprint, we selected the numerical result where the location of the tip of the rivulet, farthest from the centre of the initial position of deposition was equal to that in the experimental image.
Having matched this single point, the rest of the morphology agrees well between experiments and simulations. However, the rivulets in the numerical computations are slightly wider
than in the experiments, and the numerical time required to reach the
footprint shown in the last snapshot is significantly shorter than the
experimental time with $t_{\mathrm{num}} \simeq \frac{1}{2}
t_{\mathrm{exp}}$. In fact, the time lag between experiments and
computations increases monotonically as the rivulets grow.

\begin{figure*}[ht!]
\centering
\includegraphics[width=0.5\textwidth]{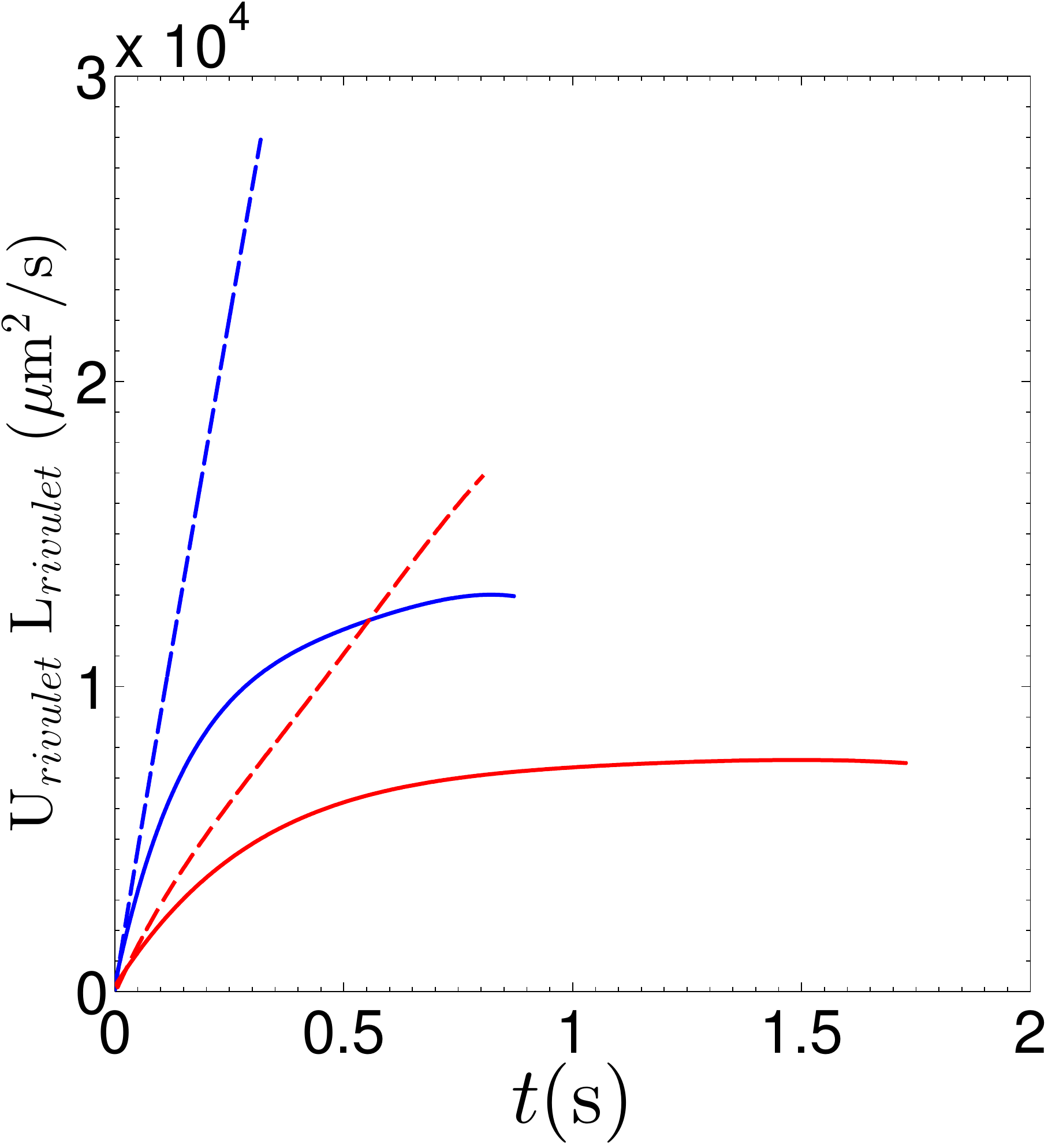}\\
\caption{Variation of $U_{\mathrm{rivulet}}\cdot L_{\mathrm{rivulet}}$ with
  time, while rivulets spread from one end of the the straight edge of the pixel boundary to another in Fig. \ref{fig:filaments_0p5Deg}(c) and \ref{fig:filaments_11Deg}(b). Solid lines denote experimental measurements, while dashed
  lines denote numerical computations performed using the experimental
  spreading laws. Blue lines are for $\theta_A < 1^\circ$ and red
  lines are for $\theta_A = 11^\circ$.}
\label{fig:simpleModel}
\end{figure*}

We believe that these differences in width and timescale arise because of increasing viscous dissipation in the
experimental rivulets as they grow; an effect that  is not included in the
model. The difference between experiments and model is highlighted in
Fig. \ref{fig:simpleModel}, where the product of the length of the
rivulet, $L_{\mathrm{rivulet}}$, and the speed of its tip,
$U_{\mathrm{rivulet}}$, measured while rivulets spread along the straight edge of the pixel boundary (see Fig. \ref{fig:filaments_0p5Deg}(c) and \ref{fig:filaments_11Deg}(b)) is shown as a function of time. 
In the model, $U_{\mathrm{rivulet}} \cdot L_{\mathrm{rivulet}}$ is
proportional to time, which means that $U_{\mathrm{rivulet}}$ is
constant. Rivulet growth is driven solely by capillary forces, and a
steady state is quickly established because of the fixed geometry and
constant wettability of the pixel. Hence, the dynamic contact angle at
the tip of the rivulet reaches a constant value, which leads to a
constant propagation speed of the tip.

 A proportional increase of $U_{\mathrm{rivulet}} \cdot L_{\mathrm{rivulet}}$
 with time is also observed in the experiments, but
 only for very early times, i.e. $t\le0.1$~s.
 $U_{\mathrm{rivulet}}$ decreases as $L_{\mathrm{rivulet}}$
 increases and $U_{\mathrm{rivulet}} \cdot L_{\mathrm{rivulet}}$ tends to a
 constant value, which is consistent with a balance between 
 the pressure gradient driving the flow in the rivulet and viscous stresses:
\begin{equation}
\frac {\Delta P}{L_{\mathrm{rivulet}}} \, \sim \, \mu \, \frac
      {U_{\mathrm{rivulet}}}{H^{2}}
\label{eq:scale1}
\end{equation}
where $\Delta P$ is the constant pressure jump across the free
surface of the rivulet at its tip, and $H$ indicates the height of
the rivulet, which is approximately equal to the height of the side wall of the pixel.  Provided that the tip geometry (and hence
the pressure drop) remains fixed, Eq. (\ref{eq:scale1})
indicates that $U_{\mathrm{rivulet}} \cdot L_{\mathrm{rivulet}}$ is
constant in this steady state regime.
 
The comparison between experiments and numerical results, obtained
using experimental spreading laws, has shown that the spreading of the
droplet in a pixel can be predicted quantitatively at short times
before viscous resistance along the rivulet becomes
significant and slows the experimental evolution. Despite these differences in timescale, the model reproduces the entire morphological
evolution until the coalescence of the rivulet tips. In other words, the correct fluid morphologies are predicted, but at the wrong times.

If measurements of the experimental spreading law are not available, then a
fit to the
Cox--Voinov law introduced in \S \ref{sec:spreading} could be used instead.
The derivation of the law from microscopic considerations
includes viscous dissipation but only very close to the contact
line \cite{cox1986dynamics,voinov}. Assuming that the spreading-law
data are not available, the velocity scale $U_1$  (see
Eq. \ref{eq:spreadinglaw}) is determined by measuring the
average slope at early times of the experimental curves in Fig.
\ref{fig:simpleModel}, when viscous resistance along the rivulet is
negligible. Note that a direct fit to the experimental spreading law in the
capillary-viscous spreading regime gave similar values of $U_1$ to
within 10\%.

\begin{figure*}[ht!]
\centering
\includegraphics[trim=4cm 14cm 4cm 2.5cm, clip=true, width=0.8\textwidth]{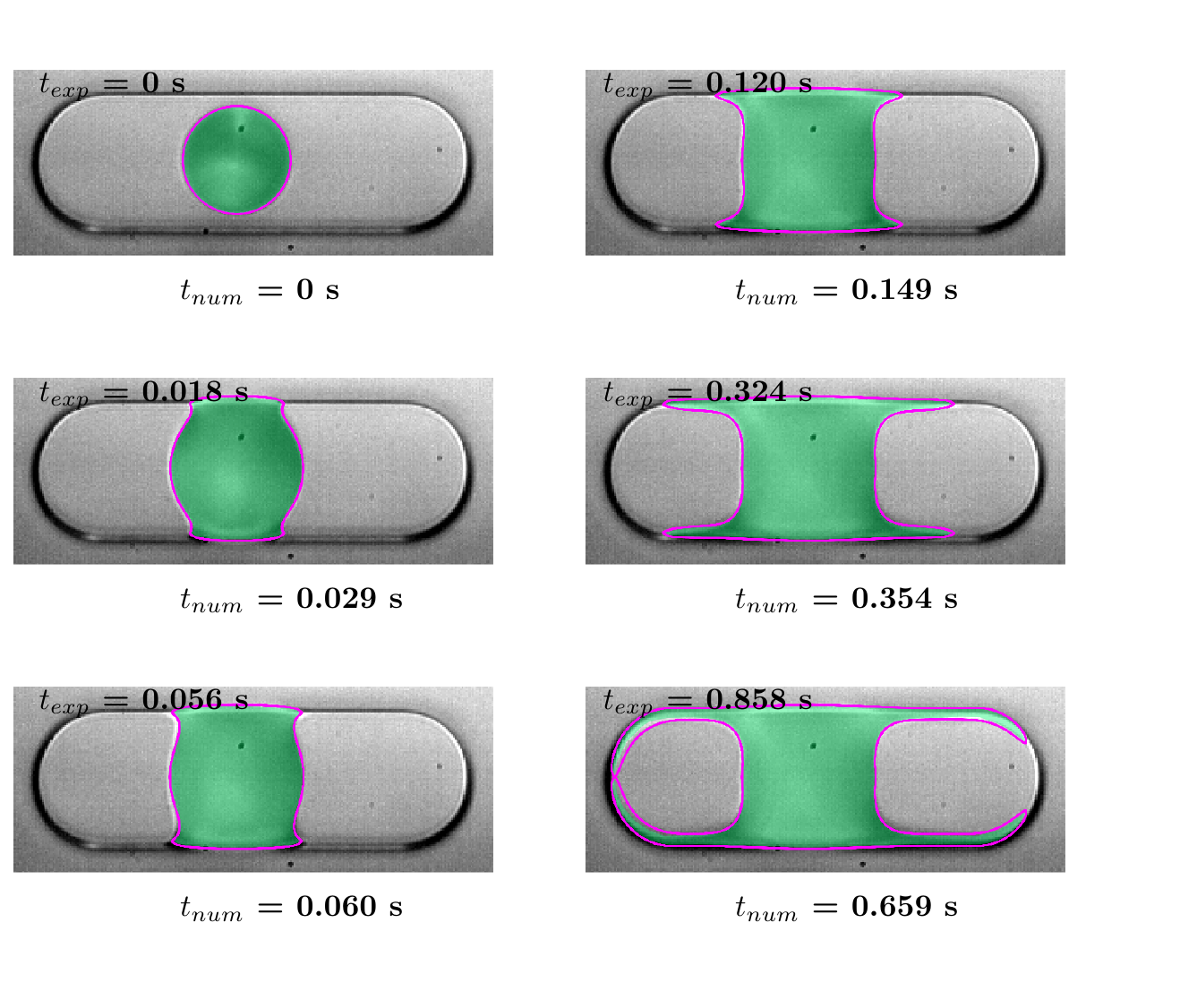}\\
\caption{Comparison between numerical calculations and experimental
  measurements for a series of snapshots of a spreading droplet
  deposited inside a pixel  with $\theta_A = 11^{\circ}$. Green shaded
  regions correspond to the experimentally measured footprint. The
  magenta lines show the outline of the droplet footprint calculated
  numerically using the Cox--Voinov spreading law presented in Fig.
  \ref{fig:fitted_spreadingLaw}(b).}
\label{fig:comparison_11}
\end{figure*}

\begin{figure*}[ht!]
\centering
\includegraphics[trim=4cm 14cm 4cm 2.5cm, clip=true, width=0.8\textwidth]{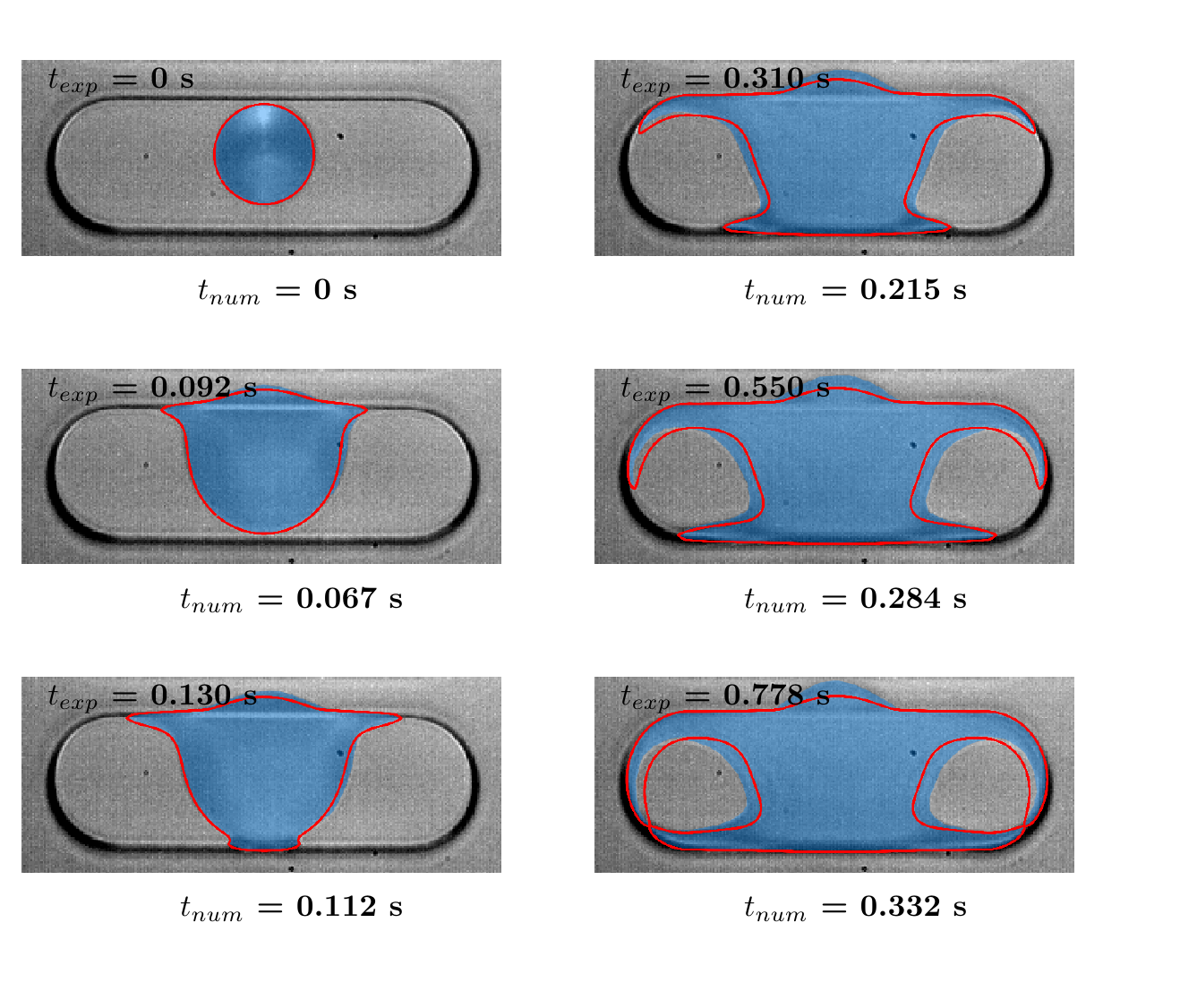}\\
\caption{Comparison between numerical calculations and experimental
  measurements for a series of snapshots of a spreading droplet
  deposited inside a pixel  with $\theta_A < 1^{\circ}$. Blue shaded
  regions correspond to the experimentally measured footprint
  (\textbf{$\theta_A < 1^{\circ}$}). The red lines show the outline of
  the droplet footprint calculated numerically using Cox--Voinov
  spreading law presented in Fig. \ref{fig:fitted_spreadingLaw}(a).}
\label{fig:comparison_0p51}
\end{figure*}

Comparisons between model predictions, using the Cox--Voinov spreading law,
and experiments are shown in Fig. \ref{fig:comparison_11} and
Fig. \ref{fig:comparison_0p51} for $\theta_A = 11^{\circ}$ and less
than 1$^{\circ}$, respectively.  Not only does the model predict the
final footprint shapes before reconnection, but it also captures the evolution of the
spreading footprint for different initial positions of the droplets.
Again, as long as the length of the rivulets remains small, the numerical and experimental
spreading time scales are in good agreement.

On the substrate with $\theta_A = 11^{\circ}$, the shape of the
footprint predicted from the model, when rivulets intersect at one end
of the pixel, matches very well with the experimental measurement
(Fig. \ref{fig:comparison_11}). In contrast, on the substrate with
$\theta_A < 1^{\circ}$ using the Cox--Voinov
spreading law leads to under-prediction of the spreading where dynamic contact angles ($\theta$) are small ($\leq
3^{\circ}$), i.e. in the middle of the pixel and in the overspill
region outside it (Fig. \ref{fig:comparison_0p51}). This is entirely
consistent with the observations in \S \ref{sec:spreading} that the
fitted Cox--Voinov law predicts lower spreading velocities than the
experiments at low contact angles.

\subsection{Effect of corner geometry}

\begin{figure*}
\begin{subfigure}{\linewidth}
\centering
\includegraphics[width=0.82\textwidth]{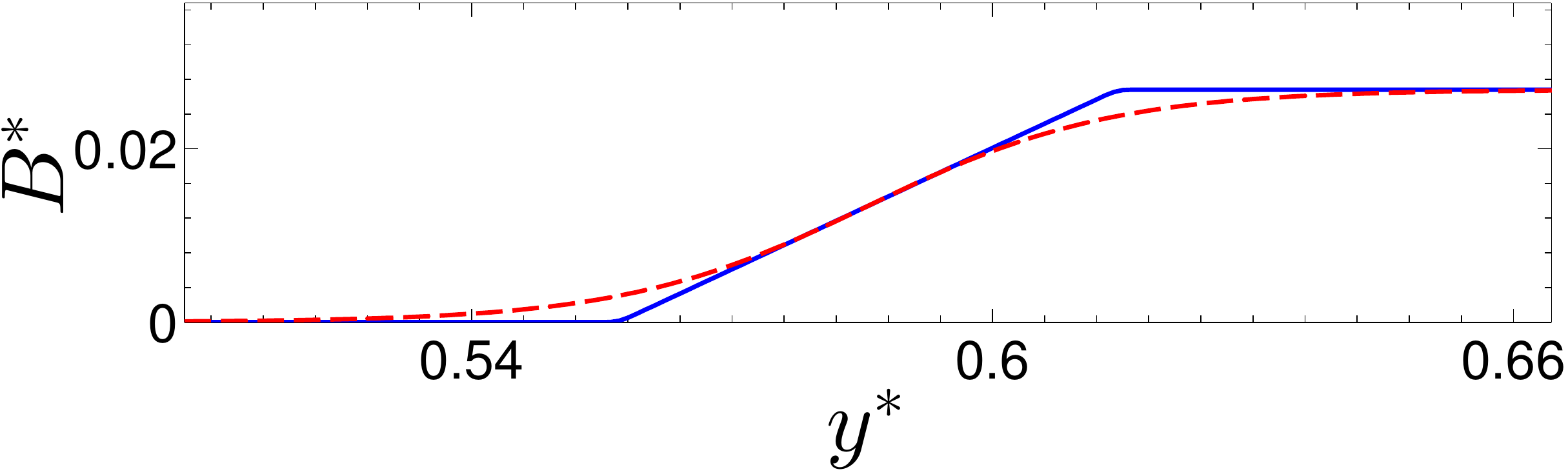}
\caption{}
\label{fig:Side_wall_Sensitive}
\end{subfigure}\\[1ex]
\begin{subfigure}{\linewidth}
\centering
\includegraphics[width=0.78\textwidth]{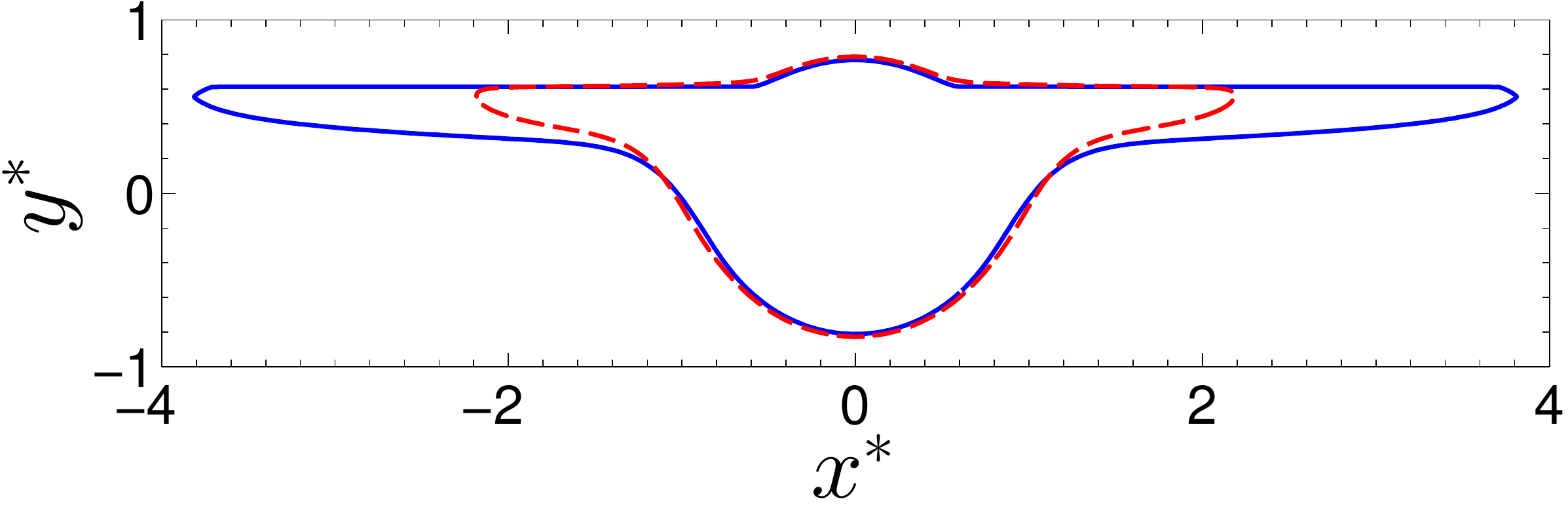}
\caption{}
\label{fig:theta_max_corner_sensitivity}
\end{subfigure}\\[1ex]
\begin{subfigure}{\linewidth}
\centering
\includegraphics[width=0.5\textwidth]{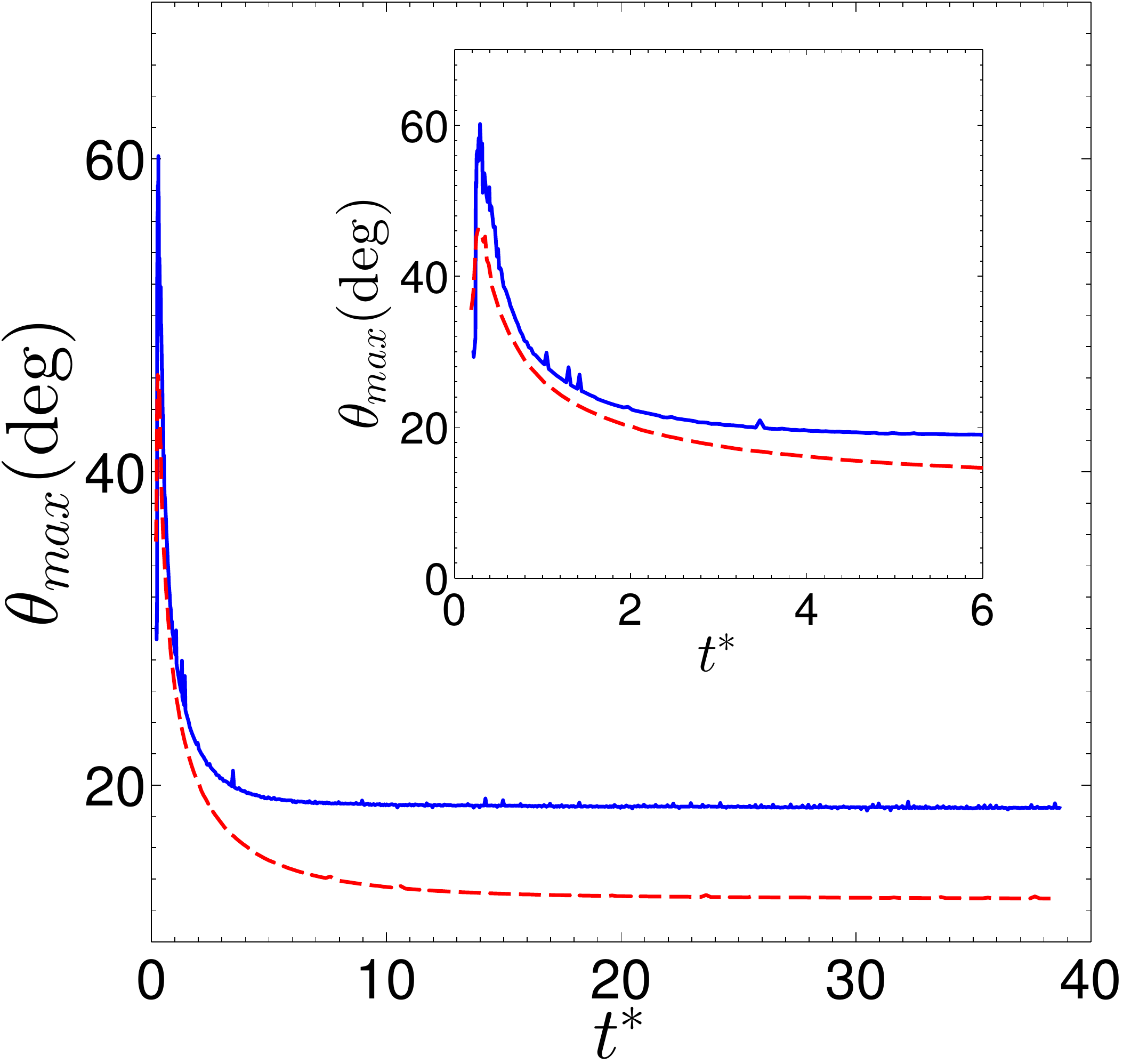}
\caption{}
\label{fig:Side_wall_Sensitive_Result}
\end{subfigure}
\caption{Spreading in corners of different sharpness. (a) Geometry of the corner: (blue) linear segments connected by circular arcs of radius $0.1 D^*/sin(25^{\circ})$, where $D^*$ is the dimensionless height of the side wall; (red) tanh function. (b) Evolution of the
  maximum dynamic contact angle ($\theta_{max})$ of the morphology. (\emph{Inset plot}) Evolution of the $\theta_{max}$ of
  the droplet just before and after the contact line first reaches the topography gradient. (c) Footprint of the spreading liquid droplet at time
  $t^*= 40$ after deposition near the topography gradient.  } \label{figcombined}
\end{figure*}

We investigated the effect of the sharpness of the corner on the speed
of the rivulet tip along the wedge using numerical simulations. A
droplet, was placed on a flat surface close to a rectilinear side wall
much wider than the drop and with a similar height profile to the
pixels investigated above ($D=1.3$~$\mu$m,
$\psi$=155$^{\circ}$). We chose two different corner shapes shown in  Fig. \ref{figcombined}(a). The sharp corners (blue line) were modelled by circular arcs of radius $0.1 D^*/sin(25^{\circ})$, where $D^*$ is the dimensionless height of the pixel side wall. Smooth corners (red line) were achieved by modelling the side wall with a hyperbolic tangent profile. The wettability of the substrate was uniform with $\theta_A =$
5$^{\circ}$ and the centre of the droplet was placed 27~$\mu$m from the bottom edge of the side wall.
As shown in Fig. \ref{figcombined}(c), the spreading of the
droplet near the wall is strongly influenced by the smoothness of the
corner. At $t^* = t \cdot U_{1}/R_{c}=40$, rivulets
have grown along the sharp corner to approximately three times the
length reached in the smooth corner. 
These differences stem from the high dynamic contact angles reached in
the convex corner, which in turn translate into locally enhanced
contact line velocities. Away from the corner, however, the
bulk of the droplet spreads in a similar manner in both geometries. The maximum values of the contact angle
reached in the simulations of spreading along sharp and smooth corners
are shown against time in Fig.
\ref{figcombined}(b). $\theta_{max}$ first increases
to a maximum value which occurs when the contact line first reaches
the corner and which is larger when the corner is sharp. This is
followed by a decrease to a constant value which is also higher in the
case of the sharp corner. This is because the term  $\vec{n}\cdot
\nabla B^*(x^*,y^*)$ in Eq. \ref{eq:theta} is larger when the corner is
sharp, so that the tip of the rivulet exhibits a larger effective contact angle
than in the case of the smooth corner.  Thus, variations
in the shape of the corner can be used to control the flow rate of the
liquid through the rivulets, and the dependence of rivulet growth
on corner sharpness can be exploited to inform the design of lab on
chip devices where transportation of small amount of liquid is
required.

\section{Conclusion}
\label{sec:conclusion}

In this paper we have presented a novel
experimental system that allows
detailed studies of the time evolution of single droplet spreading
within pixel wells, which has
applications to the ink-jet printing of POLED displays. For
sufficiently wetting substrates, we find that interaction with the
side walls enhances spreading via the mechanism of capillary rise
in sharp corners quantified by \citeauthor{concusFinn}
\cite{concusFinn}. As the contact angle increases, meaning that the
surface is less wetting, the capillary rise diminishes and eventually
above a threshold value, predicted by the Concus \& Finn formula, it
ceases altogether and the droplets are spread over a much smaller area
of the pixel. In contrast, a droplet on a raised podium is
restricted in its spreading by the bounding wall even at low contact
angles.

 A generalisation of the thin-film model developed by \citeauthor{alice}
\cite{alice} is shown to correctly reproduce the morphological
development of the droplet. The model assumes a quasi-static evolution
so that the liquid immediately adopts a surface of constant curvature
in response to changes in the contact line, which evolves according to
a spreading law relating the contact line speed to the contact
angle. The absence of viscous effects means that the timescale for
evolution of the morphology is incorrect apart from at small times
after deposition, but the morphology itself is in excellent agreement
with the experiments if an experimentally measured spreading law is used.
Moreover, the classic Cox--Voinov law \cite{cox1986dynamics, voinov}
provides a reasonable approximation of the experimental spreading law
apart from at the very smallest contact angles. Hence, the droplet
evolution, again without the correct timescale,
is adequately predicted using the Cox--Voinov law after using
a simple fit to determine the unknown velocity scale.

 Thus, we conclude that the simple model can be used
effectively to predict evolution of liquid morphology over complex topographies.

\begin{acknowledgments}
The authors would like to thank Malcolm Walker and Carl Dixon for
technical support while building the experimental facility.  We
acknowledge funding from the University of Manchester and Cambridge
Display Technology (CDT).  We are also very grateful to CDT for
providing required substrates for this study.
\end{acknowledgments}

\bibliography{article_revision_10}
BibTeX.

\end{document}